\documentclass[journal]{IEEEtran}

\ifCLASSINFOpdf
  \usepackage[pdftex]{graphicx}
	\graphicspath{ {images/} }

\else

\fi

\usepackage{amsmath}

\usepackage{url}
\usepackage{cite}
\usepackage{blindtext}
\usepackage{caption}
\usepackage{subcaption}
\usepackage[dvipsnames]{xcolor}
\usepackage{amssymb}
\usepackage{enumitem}
\usepackage{lipsum}
\usepackage{mathtools}
\usepackage{mathtools, cuted}
\usepackage{lipsum, color}
\usepackage[margin=15mm]{geometry}
\usepackage{multicol}
\usepackage{float}
\usepackage{subcaption}

\floatstyle{plain}
\newfloat{twocolequfloat}{b}{zzz}
\floatname{twocolequfloat}{Equation}

\begin{document}

\title{An Interference-Free Filter-Bank Multicarrier System Applicable for MIMO Channels}

\author{\IEEEauthorblockN{Mohammad Towliat}
\vspace{10pt}

\IEEEauthorblockA{\textit{Department of Electrical and Computer Engineering} \\
\vspace{2pt}
\textit{University of Delaware}\\
\vspace{2pt}
\textit{Newark, DE, USA} \\
\vspace{5pt}
mtowliat@udel.edu}
\vspace{-5pt}}

\maketitle

% As a general rule, do not put math, special symbols or citations
% in the abstract or keywords.
\begin{abstract}
In filter-bank multicarrier (FBMC) systems the intrinsic interference is the major challenge to achieve a full gain of diversity over multi-input multi-output (MIMO) channels. In this paper, we develop a novel multicarrier system called FBMC offset upper-lower polyphase network (FBMC/OULP) in which, to eliminate the intrinsic interference, the complex-valued symbols are alternatively transmitted via upper and lower half of polyphase network branches with an offset time. The symbol density of the FBMC/OULP system is equal to one complex-valued symbol in time-frequency lattice. Also, for transmission over frequency selective channels, a minimum mean square error (MMSE) estimator is employed at the receiver of the FBMC/OULP system to eliminate the interference caused by the frequency selectivity of the channel. The proposed scheme mitigates the produced interference between symbols in the upper and lower polyphase branches, based on the circular convolutional property. As a result of using complex-valued symbols and diminishing the interference, the full diversity gain of the orthogonal space-time block codes (OSTBC) can be achieved in MIMO channels by a low complex maximum likelihood (ML) detector. In comparison with the orthogonal frequency division multiplexing (OFDM) system, simulation results indicate that the proposed system achieves a superior performance in fast multi-path fading channels and a competitive performance in slow multi-path fading channels.
\end{abstract}

% Note that keywords are not normally used for peerreview papers.

% For peer review papers, you can put extra information on the cover
% page as needed:
% \ifCLASSOPTIONpeerreview
% \begin{center} \bfseries EDICS Category: 3-BBND \end{center}
% \fi
%
% For peerreview papers, this IEEEtran command inserts a page break and
% creates the second title. It will be ignored for other modes.
\IEEEpeerreviewmaketitle

\section{Introduction}
% The very first letter is a 2 line initial drop letter followed
% by the rest of the first word in caps.
% 
% form to use if the first word consists of a single letter:
% \IEEEPARstart{A}{demo} file is ....
% 
% form to use if you need the single drop letter followed by
% normal text (unknown if ever used by the IEEE):
% \IEEEPARstart{A}{}demo file is ....
% 
% Some journals put the first two words in caps:
% \IEEEPARstart{T}{his demo} file is ....
% 
% Here we have the typical use of a "T" for an initial drop letter
% and "HIS" in caps to complete the first word.
\IEEEPARstart{M}{ulticarrier} systems are known as the appropriate platforms to overcome the inter symbol interference (ISI) because of using easier equalization, in comparison with single-carrier systems \cite{R1}. Orthogonal frequency division multiplexing (OFDM) is the most prevalent multicarrier system, in which using a cyclic prefix (CP) with the length equal to the channel impulse response (CIR) duration, guarantees zero ISI and timing offset errors \cite{R2}. However, applying a rectangular pulse shape causes a large spectral side lobes in each subchannel that makes the OFDM very sensitive to the carrier frequency offsets (CFO) \cite{R3, R4}. In contrary to the OFDM, Chang \cite{R5} and Saltzberg \cite{R6} developed a filter-bank multicarrier (FBMC) structure using an appropriate prototype filter shape instead of the rectangular one. In \cite{R6}, the authors proposed an FBMC system associating with the offset quadrature amplitude modulation (FBMC/OQAM) in which just real-valued symbols are transmitted. In the FBMC/OQAM scheme, time interval of transmitted symbol is equal to the half of the minimum orthogonality interval of the prototype filter in the time domain; the frequency interval of transmitted symbol is also equal to the minimum orthogonality interval of the prototype filter in the frequency domain. Thus, the symbol density of the FBMC/OQAM scheme becomes two real-valued symbols in time-frequency lattice, which is equivalent to one complex-valued symbol. 

Due to using the filter-bank, the FBMC/OQAM system does not require to employ a CP during the transmission process. This feature leads to a higher bandwidth efficiency of the FBMC/OQAM compared to the OFDM, especially in the presence of channels with long CIRs. Also, it is remarkable that the FBMC-based systems are less sensitive to the CFO, particularly when the CFO becomes very intense \cite{R7, R8, R9}.

	On the other hand, in single-input single-output (SISO) frequency selective channels, the pure imaginary interference can be removed at the receiver side of the FBMC/OQAM, after eliminating the effect of channel by one-tap zero forcing (ZF) or minimum mean square error (MMSE) equalizers. However, in multi-input multi-output (MIMO) channels, it is not possible to achieve full diversity gain of space-time block codes (STBC) in the FBMC/OQAM system \cite{R10}. Many attempts have been made to tackle this issue in the FBMC/OQAM. Per-subchannel MMSE equalizers have been developed in \cite{R11, R12, R13, R14, R15} to extract the desired symbols. The idea is based on either the direct estimation of the desired symbols or estimation of the symbols in the vicinity to remove their effects on desired ones. Also, in \cite{R16} a combined MMSE and iterative maximum likelihood (ML) procedure has been used to improve the interference cancellation over QAM data symbols coded as Alamouti orthogonal STBC (OSTBC).
	
Furthermore, joint precoding and MMSE estimation methods have been proposed to overcome the interference in the FBMC/OQAM in \cite{R17, R18, R19, R20, R21, R22}. Decision feedback equalizers (DFE) have been developed in \cite{R23, R24, R25, R26, R27, R28} for FBMC systems, as well. A parallel equalization for transmission over intense frequency selective channels has been utilized in \cite{R29}, where one-tap equalizer per each subchannel becomes a special case when the number of subchannels is very high. Interference cancellation methods, including oversampling when system uses QAM symbols, code division multiple access (CDMA) and Viterbi-based technique have been designed in \cite{R30, R31, R32}, respectively, for the FBMC systems.
In \cite{R33}, the authors have developed a method based on signal to leakage pulse noise ratio (SLNR) and signal to interference pulse noise ratio (SINR) maximization to decrease the BER. A full rate realization method of the OSTBC has been proposed in \cite{R34} for FBMC systems to remove the imaginary interference in the presence of the channel effects. The main defect of this method is the diversity loss of the MIMO channel which leads to the performance degradation.

It is remarkable that, similar to OFDM, a CP can be utilized in the FBMC systems to eliminate the interferences. In this regard, an innovative structure called fast Fourier transform-FBMC (FFT-FBMC) has been proposed in \cite{R35} to mitigate the inherent interference by using a separated OFDM tendency in each subchannel. This system needs to insert a short CP to each block of QAM transmitted symbols to get over the intrinsic interference caused by filter-bank structure. As a result, the ML detection of Alamouti OSTBC can be implemented along with the proposed FFT-FBMC system.

In this article, to overcome the intrinsic interference, we develop a new FBMC system in which the complex-valued symbols are alternatively loaded on the upper and lower branches of polyphase network with an offset time. The system is called FBMC offset upper-lower polyphase network (FBMC/OULP) where the offset time is equal to the half of the minimum orthogonality time duration that is exactly the same as the offset time in the FBMC/OQAM system. To eliminate the interference between upper and lower branches, originated form the frequency selectivity of channel, an interference eliminator (IE) is designed at the receiver of the proposed FBMC/OULP system, based on circular overlap between desired and interfered parts. Since only a few number of symbols (related to the CIR length) make overlap and produce interference, the IE is efficiently able to mitigate the interference. Thus, instead of estimating the interference of all subchannels, which is performed in \cite{R14, R15, R16}, the interference of a few numbers of symbols are needed to be estimated in the FBMC/OULP system.  Moreover, due to not using a CP, in comparison with the OFDM and FFT-FBMC systems, the proposed schemes is more bandwidth efficient.

The remaining parts of this paper are organized as follows. The primer FBMC/OQAM system is introduced in the next section. Section III includes step-by-step development of the proposed FBMC/OULP system for SISO channels. Section IV extends the developed SISO FBMC/OULP system for the MIMO channels. Simulation results are given in Section V and finally, section VI contains the article conclusions.  

\section{FBMC/OQAM System Model}
Let assume that ${{d}_{k,l}}$ is the data symbol of $k\text{th}$ time slot and the $l\text{th}$ subchannel, the transmitted signal of the FBMC/OQAM system with $L$ subchannels can be written as
\begin{equation}
s(m)=\sum\limits_{k}^{{}}{\sum\limits_{l=0}^{L-1}{{{d}_{k,l}}}\,\,}{{f}_{k,l}}(m),
\label{1}
\end{equation}
where ${{f}_{k,l}}(m)\triangleq f(m-k\,\Delta T)\,\,{{e}^{j2\pi \,l\,\Delta F\,m}}{{e}^{j\pi (l+k)/2}}$ is the time and frequency shifted version of a real-valued prototype filter $f(m)$. Although it is not possible to design a prototype filter to satisfy the complex-orthogonality condition when $\Delta T\times \Delta F=1$ \cite{R36}, for $\Delta T\times \,\Delta F=0.5$, one can design the $f(m)$ filter in order to satisfy the following real-orthogonality condition \cite{R35}  
\begin{equation}
\text{Re}\left\{ \sum\limits_{m=-\infty }^{\infty }{{{f}_{k,l}}(m)\,f_{{k}',{l}'}^{*}(m)} \right\}\,={{\delta }_{\kappa,\ell }},
\label{2}
\end{equation}
where ${{\delta }_{k,\ell }}$ is the Dirac delta function and $\kappa \triangleq {k}'-k$ and $\ell \triangleq {l}'-l$ are the time and frequency spaces between filters’ indices. In the FBMC/OQAM, ${{d}_{k,l}}$ is considered as a real-valued symbol when the time and frequency shifts are set to $\Delta T=L/2$ and $\Delta F=1/L$. Thus, for the real-valued symbols, the data symbol density in time-frequency lattice becomes $\beta \triangleq 1/(\Delta T\times \,\Delta F)=2$ which is equivalent to $\beta =1$ for complex-valued symbols. With the aim of extracting the desired symbol $\,{{d}_{{k}',{l}'}}$, by using $f_{{k}',{l}'}^{*}(m)$ as a matched-filter at the receiver, in the AWGN channel, the output becomes \cite{R35}
\begin{equation}
\begin{array}{*{20}{l}}
{{y_{k',l'}} = \sum\limits_{m =  - \infty }^\infty  {[s(m) + w(m)]f_{k',l'}^*(m)}  = }\\
{{d_{k',l'}} + \sum\limits_{k \ne k'}^{} {\sum\limits_{\begin{array}{*{20}{c}}
{l = 0}\\
{l \ne l'}
\end{array}}^{L - 1} {{d_{k,l}}\sum\limits_{m =  - \infty }^\infty  {{f_{k,l}}(m)\,f_{k',l'}^*(m)} } }  + {\omega _{k',l'}},}
\end{array}
\label{3}
\end{equation}
where $w\,(m)$ is the zero mean AWGN noise with power of ${{N}_{0}}$ and ${{\omega }_{{k}',{l}'}}\triangleq $ $\sum\nolimits_{m=-\infty }^{\infty }{w\,(m)}$ $f_{{k}',{l}'}^{*}(m)$. Note that by considering \eqref{2}, the interference caused by other symbols (${{d}_{k,l}}$ when $k\ne {k}'$ or $l\ne {l}'$) is pure imaginary and the real-valued desired symbol ${{d}_{{k}',{l}'}}$ can be detected by taking the real part of the matched-filter output. 

However, in the case of multi-path and time-variant channels with impulse response ${{h}_{{k}',m}}$ (for $m=0,1,\ldots {{L}_{\text{c}}}-1$, ${{h}_{{k}',m}}$ presents the CIR with length ${{L}_{\text{c}}}$ at the ${k}'\text{th}$ time slot), the symbol ${{d}_{{k}',{l}'}}$ can be detected by real-taking operator after dividing the matched-filter output to ${{H}_{{k}',{l}'}}\triangleq $ $\sum\nolimits_{m=0}^{{{L}_{\text{c}}}-1}{{{h}_{{k}',m}}\,{{e}^{-j2\pi {l}'\,m/L}}}$, where ${{H}_{{k}',{l}'}}$ is the ${l}'\text{th}$ element of $L\text{-point}$ FFT of ${{h}_{{k}',m}}$ \cite{R37}.

When the STBC is employed in the MIMO communication system, it is necessary to transmit complex-valued symbols in order to achieve the full diversity gain by using the ML detection scheme with a low complexity. However, due to using real-valued symbols, the FBMC/OQAM scheme cannot achieve the full diversity gain in the MIMO communication system. In the following, we propose a novel FBMC system that overcomes the interference when complex-valued symbols are transmitted. In the proposed system, called FBMC/OULP, after removing the interference caused by the frequency selectivity of channel, a simple ML method can be used to detect the OSTBC. In order to follow up the designing procedure easier, first, we develop the FBMC/OULP system for SISO channels, and then it will be extended for MIMO channels.

\section{The FBMC/OULP System for SISO channels}
In this section, the structure of the FBMC/OULP system is derived for SISO channels in two steps. At the first step, the system structure is developed for frequency flat channel (AWGN channel) and then the first proposed system is modified for frequency selective channels.

\subsection{In frequency flat channels}
In the FBMC/OULP, ${{f}_{k,l}}(m)$ is chosen similar to that of the FBMC/OQAM and ${{d}_{k,l}}$ is generally a complex-valued symbol. Consequently, as \eqref{3} illustrates, in the case of frequency flat channel, ${{y}_{{k}',{l}'}}$ (the output of matched-filter at the ${k}'\text{th}$ time slot and ${l}'\text{th}$ subchannel) contains the desired symbol ${{d}_{{k}',{l}'}}$ interfered by its surrounding symbols in time and frequency domains according to $\kappa={k}'-k$ and $\ell ={l}'-l$. For more facility, let’s define $\xi _{\kappa,\ell }^{{{k}'}}$ as the interference extent of the specific symbol ${{d}_{k,l}}$ into the desired symbol ${{d}_{{k}',{l}'}}$ 
\begin{equation}
\begin{array}{l}
\xi _{\kappa,\ell }^{k'} \buildrel \Delta \over = {\mkern 1mu} \sum\limits_{m =  - \infty }^\infty  {{\mkern 1mu} {f_{k,l}}(m){\mkern 1mu} f_{k',l'}^*(m)} \\
{\rm{      }} \,\,\,\,\,\,\,\,\, =\left\{ {\sum\limits_{m =  - \infty }^\infty  {f(m + \kappa \frac{L}{2})f(m){\mkern 1mu} {e^{ - j2\pi \ell \,\,m/L}}} } \right\}\\
\,\,\,\,\,\,\,\,\,\,\,\,\,\,\,{\mkern 1mu}  \times {e^{ - j\pi \,\ell k'}}{\mkern 1mu} {e^{ - j\frac{\pi }{2}(\kappa + \ell )}}.
\end{array}
\label{4}
\end{equation}
As it can be seen, $\xi _{\kappa,\ell }^{{{k}'}}$ is a function of $\kappa$, $\ell $ and also ${k}'$ (it is clear that $\xi _{\kappa,\ell }^{{{k}'}}$ differs according whether ${k}'$ is an even or an odd number). Since in \eqref{4}, $-\infty \le \kappa<\infty $ and $-L+1\le \ell \le L-1$, \eqref{3} can be rewritten as
\begin{equation}
{{y}_{{k}',{l}'}}={{d}_{{k}',{l}'}}+\sum\limits_{\begin{smallmatrix} 
\kappa=-\infty  \\ 
 \kappa \ne 0 
\end{smallmatrix}}^{\infty }{\,\sum\limits_{\begin{smallmatrix} 
 \ell =-L+1 \\ 
 \ell \ne 0 
\end{smallmatrix}}^{L-1}{\xi _{\kappa,\ell }^{{{k}'}}\,{{d}_{{k}'-\kappa,{l}'-\ell }}+{{\omega }_{{k}',{l}'}}}}.
\label{5}
\end{equation}

As is obvious, the quantity of $\xi _{\kappa,\ell }^{{{k}'}}$ presents the contribution of ${{d}_{k,\,l}}$ to the ${y}_{{k}',{l}'}$ so that $\xi _{0,0}^{{{k}'}}=1$ is assigned to the desired symbol and $\xi _{\kappa,\ell }^{{{k}'}}$ when $(\kappa,\ell )\ne (0,0)$,  is dedicated to the undesired interference from surrounding symbols. To have a sense about the values of $\xi _{\kappa,\ell }^{{{k}'}}$, Table I exhibits $\xi _{\kappa,\ell }^{{{k}'}}$ coefficients when $f(m)$ is isotropic orthogonal transform algorithm (IOTA) \cite{R38}, which is one of the most popular prototype filters in the FBMC systems. Note that in Table I, $\xi _{\kappa,\ell }^{{{k}'}}$ is presented for even values of ${k}'$. According to \eqref{4}, for odd values of ${k}'$, the $\xi _{\kappa,\ell }^{{{k}'}}$ quantities shown in Table I, would be multiplied by $\exp (j\pi \,\ell )$.

On the other hand, when assuming that $L$ is dividable by $4$, it leads to
\begin{equation}
\begin{array}{l}
\xi _{\kappa,\ell  \pm L}^{k'} = \left\{ {\sum\limits_{m =  - \infty }^\infty  {f(m + \kappa \frac{L}{2})f(m)\,{e^{ - j2\pi (\ell  \pm \,L)\,\,\,m/L}}} } \right\}\\
\,\,\,\,\,\,\,\,\,\,\,\,\,\,\,\,\,\,\,\,\,\,\, \times {e^{ - j\pi \,(\ell  \pm \,L)\,k'}}\,{e^{ - j\frac{\pi }{2}(\kappa + \ell  \pm \,\,L)}} = \xi _{\kappa,\ell }^{k'}.
\end{array}
\label{6}
\end{equation}
Thus, by considering this circular property of $\xi _{\kappa,\ell }^{{{k}'}}$, in a matrix formulation, \eqref{5} can be represented as
\begin{equation}
{{\bar{y}}_{{{k}'}}}=\sum\limits_{\kappa=-\infty }^{\infty }{\mathbf{Z}_{\kappa}^{{{k}'}}\,}{{\bar{d}}_{{k}'-\kappa}}+{{\bar{\omega }}_{{{k}'}}},
\label{7}
\end{equation}
where ${{\bar{y}}_{{{k}'}}}\triangleq {{[{{y}_{{k}',0}},\,{{y}_{{k}',1}},\ldots {{y}_{{k}',L-1}}]}^{T}}$, ${{\bar{d}}_{k}}\triangleq {{[{{d}_{k,0}},\,{{d}_{k,1}},\ldots {{d}_{k,L-1}}]}^{T}}$ and ${{\bar{\omega }}_{{{k}'}}}\triangleq {{[{{\omega }_{k,0}},\,{{\omega }_{k,1}},\ldots {{\omega }_{k,L-1}}]}^{T}}$; also the circular matrix $\mathbf{Z}_{\kappa}^{{{k}'}}\,$ with size $L\times L$ and components $\xi _{\kappa,\ell }^{{{k}'}}$  is defined as
\begin{equation}
{\bf{Z}}_\kappa ^{k'} = \left[ {\begin{array}{*{20}{c}}
{\xi _{\kappa ,0}^{k'}}& \cdots &{\xi _{\kappa , - \Delta }^{k'}}&{\bf{0}}&{\xi _{\kappa ,\Delta }^{k'}}& \cdots &{\xi _{\kappa ,1}^{k'}}\\
 \vdots & \searrow & \vdots & \searrow &{}& \searrow & \vdots \\
{\xi _{\kappa ,\Delta }^{k'}}&{}&{\xi _{\kappa ,0}^{k'}}&{}&{}&{}&{\xi _{\kappa ,\Delta }^{k'}}\\
{\bf{0}}&{}& \vdots & \searrow &{}&{}&{\bf{0}}\\
{\xi _{\kappa , - \Delta }^{k'}}&{}&{\xi _{\kappa ,\Delta }^{k'}}&{}&{}&{}&{\xi _{\kappa , - \Delta }^{k'}}\\
 \vdots & \searrow &{\bf{0}}& \searrow &{}& \searrow & \vdots \\
{\xi _{\kappa , - 1}^{k'}}& \cdots &{\xi _{\kappa , - \Delta }^{k'}}&{\bf{0}}&{\xi _{\kappa ,\Delta }^{k'}}& \cdots &{\xi _{\kappa ,0}^{k'}}
\end{array}} \right],
\label{8}
\end{equation}
in which $\Delta $ is a number such that for $\left| \ell  \right|>\Delta $, the validity of approximation $\xi _{\kappa,\ell }^{{{k}'}}\approx 0$ is guaranteed. Regarding Table I, we can consider $\Delta =1$, in the case of IOTA.
In addition, when using IOTA as the time-frequency well localized prototype filter, the interference through the time axis, is mainly coming from the adjacent symbols and the impacts of the other symbols are negligible \cite{R35} (also refer to Table I). Thus, \eqref{7} can be approximated as\footnote{Note that the approximation in \eqref{9} associates with more tolerance when the prototype filters are more spread in time, such as PHYDYAS prototype filter \cite{R39}.}
\begin{equation}
{{\bar{y}}_{{{k}'}}}=\sum\limits_{\kappa=-1}^{1}{\mathbf{Z}_{\kappa}^{{{k}'}}\,}{{\bar{d}}_{{k}'-\kappa}}+{{\bar{\omega }}_{{{k}'}}}.
\label{9}
\end{equation}

After taking an inverse FFT (IFFT) of both sides of \eqref{9}, it becomes 
\begin{equation}
{{\bar{Y}}_{{{k}'}}}=\sum\limits_{\kappa=-1}^{1}{\mathbf{V}_{\kappa}^{{{k}'}}{{{\bar{D}}}_{{k}'-\kappa}}\,}+{{\bar{\Omega }}_{{{k}'}}},
\label{10}
\end{equation}
where ${{\bar{Y}}_{{{k}'}}}\triangleq \mathbf{F}_{L}^{\dagger }{{\bar{y}}_{{{k}'}}}$, $\mathbf{V}_{\kappa}^{{{k}'}}\triangleq \mathbf{F}_{L}^{\dagger }\mathbf{Z}_{\kappa}^{{{k}'}}{{\mathbf{F}}_{L}}$ and ${{\bar{\Omega }}_{{{k}'}}}\triangleq \mathbf{F}_{L}^{\dagger }{{\bar{\omega }}_{{{k}'}}}$, in which ${{\mathbf{F}}_{L}}$ is the $L$-point FFT matrix with the $(a,b)\text{th}$ entry equal to $\exp (-j2\pi ab/L)$ and ${{(.)}^{\dagger }}$ indicates the conjugate transpose operator. Also, in \eqref{10} we assume that the symbol vector ${{\bar{d}}_{k}}$ is the FFT of the new version of the transmitted symbols ${{\bar{D}}_{k}}\triangleq {{[{{D}_{k,0}},\,{{D}_{k,1}},\,\ldots {{D}_{k,L-1}}]}^{T}}$ such that ${{\bar{d}}_{k}}={{\mathbf{F}}_{L}}{{\bar{D}}_{k}}$.

It is shown in Appendix A that $\mathbf{V}_{\kappa}^{{{k}'}}$ is a diagonal matrix such that $\mathbf{V}_{\kappa}^{{{k}'}}\text{=diag}(\bar{v}_{\kappa}^{{{k}'}})$ in which for $\bar{\xi }_{\kappa}^{{{k}'}}\triangleq [\xi _{\kappa,0}^{{{k}'}},\ldots \xi _{\kappa,\Delta }^{{{k}'}},0,\ldots 0,\xi _{\kappa,-\Delta }^{{{k}'}},\ldots \xi _{\kappa,-1}^{{{k}'}}]{{\,}^{T}}$ with size $L\times 1$, ${\bar{v}}_{\kappa}^{{{k}'}}$ is defined as
\begin{equation}
{\bar{v}}_{\kappa}^{{{k}'}}\triangleq \mathbf{F}_{L}^{\dagger }\bar{\xi }_{\kappa}^{{{k}'}}.
\label{11}
\end{equation}
Assuming ${\bar{v}}_{\kappa}^{{{k}'}}\triangleq {{[{v}_{\kappa,0}^{{{k}'}},\,{v}_{\kappa,1}^{{{k}'}},\ldots \,{v}_{\kappa,L-1}^{{{k}'}}]}^{\,T}}$, in the case of IOTA, the $n\text{th}$ component of ${\bar{v}}_{0}^{{{k}'}}$ yields to be
\begin{equation}
v_{0,n}^{{{k}'}}\,=1+0.8822\,\sin (2\pi \frac{n}{L}-\pi {k}'),\,\text{for}\,\,n=0,1,\ldots L-1.
\label{12}
\end{equation}
Therefore, when ${k}'$ is an even number, then ${v}_{0,n}^{{{k}'}}\,>1$ for $n=0,\ldots L/2-1$ and $0<{v}_{0,n}^{{{k}'}}\,<1$ for $n=L/2,\ldots \,L-1$; and when ${k}'$ is an odd number, it is vice versa. For example when $L=32$, Fig. 1 shows the values of ${v}_{0,n}^{{{k}'}}$ for even and odd ${k}'$.

\begin{figure}[!t]
\centering
\includegraphics [width=2in]{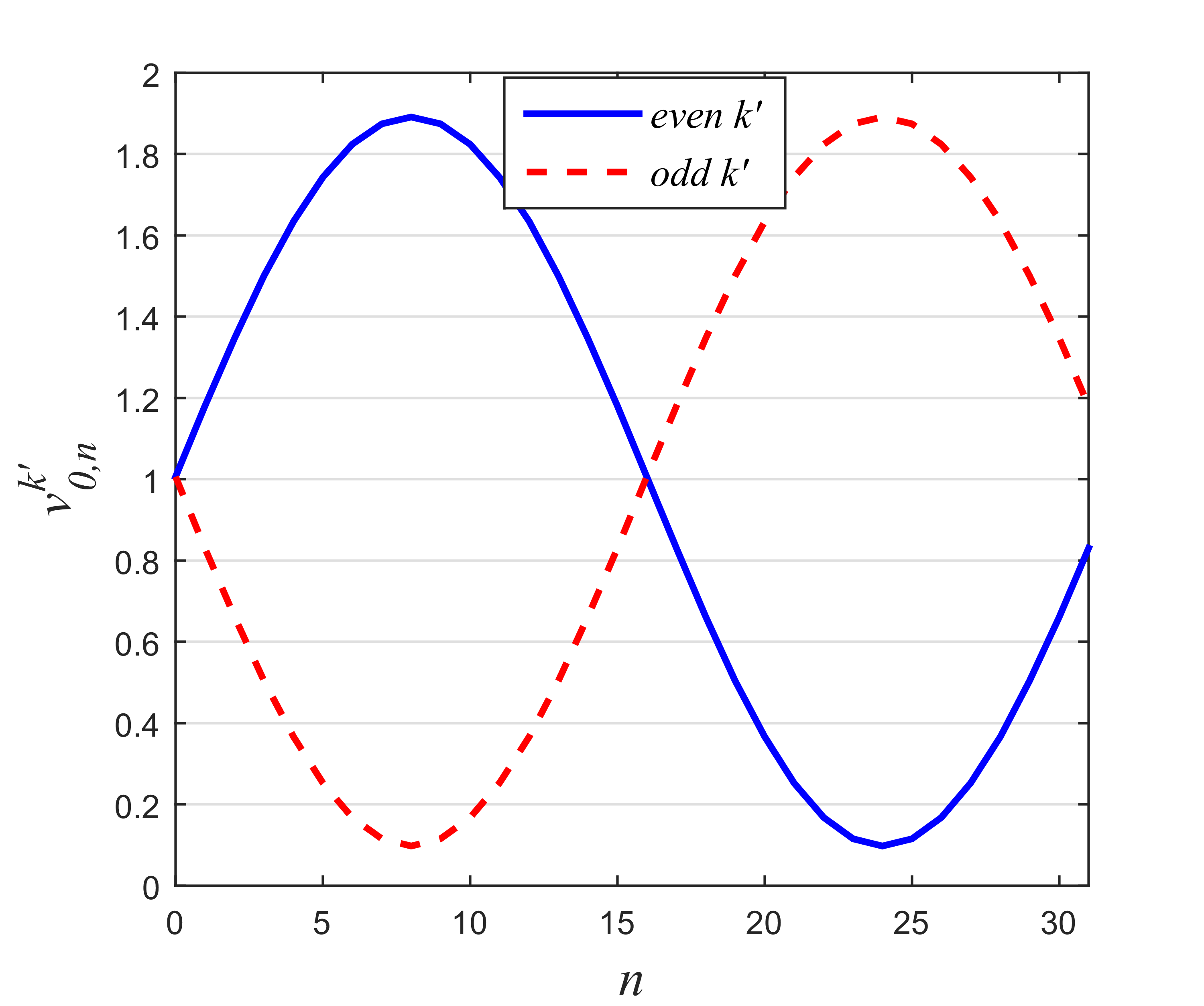}
\vspace{-3mm}
\caption{The value of $v_{0,n}^{{{k}'}}$ for even and odd ${k}'$ when $L=32$.} \vspace{-5mm}
 \label{F1}
\end{figure}

Because of diagonal structure of $\mathbf{V}_{\kappa}^{{{k}'}}$ in \eqref{10}, it is obvious that in ${{\bar{Y}}_{{{k}'}}}$, every individual desired symbol ${{D}_{{k}',n}}$ suffers from just the interference of ${{D}_{{k}'-1,n}}$ and ${{D}_{{k}'+1,n}}$. By considering \eqref{12}, Fig. 2 presents the FBMC/OULP transmission policy, in which to get rid of the interference, the complex-valued symbols are loaded on the lower half of vector ${{\bar{D}}_{k}}$ (${{D}_{k,n}}$ when $n=0,\ldots L/2-1$) and the upper half is set to zeros, when $k$ is even. In contrary, when $k$ is odd the complex-valued symbols are loaded on the upper half of vector ${{\bar{D}}_{k}}$ (${{D}_{k,n}}$ when $n=L/2,\ldots L-1$) and the lower half is set to zeros. This strategy, firstly, leads to canceling the interference of adjacent symbols, illustrated in \eqref{10}; secondly, causes the maximum transmission gain for the desired symbols ($v_{0,n}^{{{k}'}}>1$ coefficients are multiplied by desired symbols in \eqref{10}).
	
\begin{figure}[!t]
\centering
\includegraphics [width=2.in]{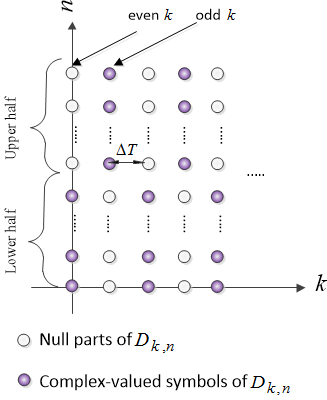}
\vspace{-0mm}
\caption{Lattice of symbol ${{D}_{k,n}}$ in the FBMC/OULP system. For even $k$, the lower half of frame is loaded and the upper half is set to zero; for odd $k$ it is vice versa.} \vspace{-3mm}
 \label{F2}
\end{figure}

\begin{table*}
\caption{The value of $\xi _{\kappa,\ell }^{{{k}'}}$ caused interference in the FBMC/OULP for the IOTA prototype filter, when $k'$ is even.}
\centering
\begin{tabular}{ |c|c|c|c|c|c|c|c|c|c| } 
\hline 
$|\xi _{\kappa,\ell }^{{{k}'}}|$ & $k=4$ & $k=3$ & $k=2$ & $k=1$ & $k=0$ & $k=-1$ & $k=-2$ & $k=-3$ & $k=-4$ \\ 
\hline
$\ell =4$  & 0	& -0.0001j	& 0	& -0.0016j	& 0 &	0.0016j	& 0	& 0.0001j &	0 \\ 
\hline
$\ell =3$ & -0.0001j &	0.0004j & -0.0015j	& 0.01j	& -0.0182j &	0.01j	& -0.0015j	& 0.0004j &	-0.0001j\\ 
\hline
$\ell =2$ & 0	&-0.0015j	& 0	& -0.038j	& 0 &	0.038j	& 0	& 0.0015j	& 0 \\ 
\hline  
$\ell =1$ & -0.0016j	& 0.01	& -0.038j	& 0.227j	& 0.4411j &	0.227j	& -0.038 j &	0.01 &	-0.0016j \\
\hline
$\ell =0$ & 0	& -0.0182j	& 0	& -0.4411j	& 1	& 0.4411j &	0 &	0.0182j &	0 \\
\hline
$\ell =-1$  & 0.0016j &	0.01 &	0.038j	& 0.227j & -0.4411j	& 0.227j &	0.038 j &	0.01 & 0.0016j \\
\hline
$\ell =-2$  & 0 &	-0.0015j &	0	& -0.038j &	0 &	0.038j &	0	& 0.0015j &	0\\ 
\hline 
$\ell =-3$  & 0.0001j	& 0.0004j	& 0.0015j &	0.01j &	0.0182j &	0.01j	 & 0.0015j &	0.0004j &	0.0001j  \\ 
\hline
$\ell =-4$ & 0	& -0.0001j &	0	& -0.0016j	& 0	& 0.0016j &	0	& 0.0001j &	0  \\ 
\hline
\end{tabular}
\label{T1}
\end{table*}

Based on the proposed symbol transmission strategy of the FBMC/OULP system, the lattice of the outputs (${{\bar{Y}}_{{{k}'}}}$) is shown in Fig. 3, where the null spaces of position $({k}',n)$ in Fig. 2 are filled with the interference of adjacent symbols and the other parts are filled with the desired symbols without any overlap. To make it more clear, \eqref{10} can be rewritten as
\begin{equation}
\begin{array}{*{20}{l}}
{{{\bar Y}_{k'}} = {\bf{V}}_0^{k'}{{\bar D}_{k'}} + \sum\limits_{\begin{array}{*{20}{c}}
{\kappa  =  - 1}\\
{\kappa  \ne 0}
\end{array}}^1 {{\bf{V}}_\kappa ^{k'}{{\bar D}_{k' - \kappa }}{\mkern 1mu} }  + {{\bar \Omega }_{k'}}}\\
{{\mkern 1mu} {\mkern 1mu} {\mkern 1mu} {\mkern 1mu} {\mkern 1mu} {\mkern 1mu} {\mkern 1mu} {\mkern 1mu} {\mkern 1mu} {\mkern 1mu} \,\,\,\,\, =  \bar Y_{k'}^{\rm{De}} + \bar Y_{k'}^{\rm{In}} + {{\bar \Omega }_{k'}}.}
\end{array}
\label{13}
\end{equation}

\begin{figure}[!t]
\centering
\includegraphics [width=2in]{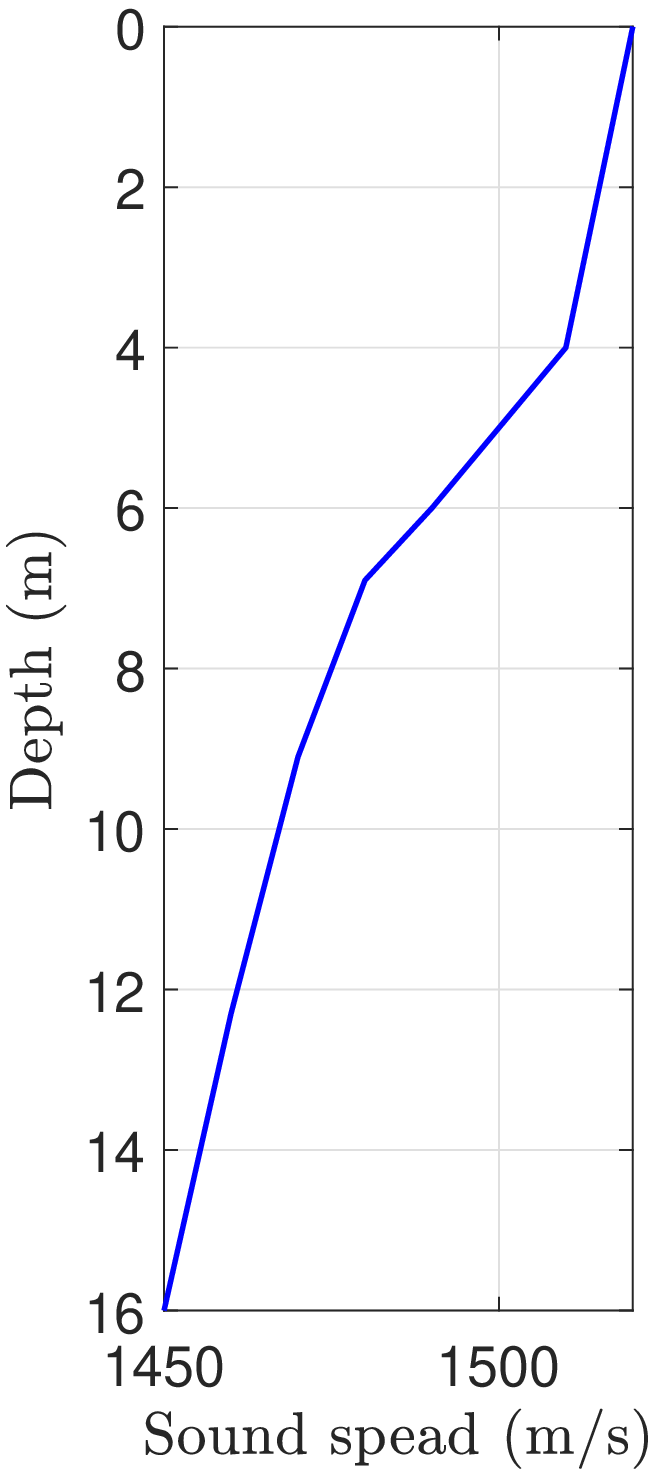}
\vspace{-0mm}
\caption{Lattice of ${{\bar{Y}}_{{{k}'}}}$ in the FBMC/OULP system when channel is frequency flat. For even ${k}'$, the lower half of the frame is desired and the upper half is interfered part; for odd ${k}'$ it is vice versa.} 
\vspace{-3mm}
 \label{F3}
\end{figure}

\noindent where $\bar{Y}_{{{k}'}}^{\rm{De}}\triangleq \mathbf{V}_{0}^{{{k}'}}{{\bar{D}}_{{{k}'}}}$ and $\bar{Y}_{{{k}'}}^{\rm{In}}\triangleq \sum\nolimits_{\begin{smallmatrix} 
 \kappa=-1 \\ 
\kappa\ne 0 
\end{smallmatrix}}^{1}{\mathbf{V}_{\kappa}^{{{k}'}}{{{\bar{D}}}_{{k}'-\kappa}}}$ are the desired and interference parts of ${{\bar{Y}}_{{{k}'}}}$. When ${k}'$ is even, since the desired symbols are loaded on the lower half of ${{\bar{D}}_{{{k}'}}}$ and its upper half is zero, the lower half of $\bar{Y}_{{{k}'}}^{\rm{De}}$ contains the desired symbols and its upper half would be zero. In contrary, since the upper half of both ${{\bar{D}}_{{k}'+1}}$ and ${{\bar{D}}_{{k}'-1}}$ are loaded with adjacent symbols and their lower halves are zeros, the upper half of $\bar{Y}_{{{k}'}}^{\rm{In}}$ contains the interference from adjacent symbols and its lower half is zero (also for odd ${k}'$ it is vice versa). Accordingly, without any overlap, $\bar{Y}_{{{k}'}}^{\rm{De}}$ and $\bar{Y}_{{{k}'}}^{\rm{In}}$ are separable which leads to detection of desired symbols at the receiver without any interference 
\begin{equation}
\begin{array}{*{20}{l}}
\begin{array}{l}
{\rm{When}}{\mkern 1mu} \,k'{\mkern 1mu} \,{\rm{is}}{\mkern 1mu} {\mkern 1mu} {\rm{even:}}\\
Y_{k',n}^{\rm{De}} = 
\end{array}\\
\begin{array}{l}
\left\{ {\begin{array}{*{20}{l}}
{{v}_{0,n}^{k'}{\mkern 1mu} {D_{k',n}},\,\,{\rm{for}}\,\,n = 0,{\mkern 1mu} {\mkern 1mu}  \ldots L/2 - 1}\\
{0,{\mkern 1mu} \,\,{\rm{for}}\,\,n = L/2,{\mkern 1mu} {\mkern 1mu}  \ldots L - 1.}
\end{array}} \right.\\
\\
Y_{k',n}^{\rm{In}} = {\mkern 1mu} {\mkern 1mu} {\mkern 1mu} {\mkern 1mu} {\mkern 1mu} 
\end{array}\\
{{\mkern 1mu} \left\{ {\begin{array}{*{20}{l}}
{0,{\mkern 1mu} {\mkern 1mu} \,\,{\rm{for}}\,\,n = 0,{\mkern 1mu} {\mkern 1mu}  \ldots L/2 - 1}\\
{{v}_{ + 1,n}^{k'}{\mkern 1mu} {D_{k' - 1,n}} + {v}_{ - 1,n}^{k'}{\mkern 1mu} {D_{k' + 1,n}},{\mkern 1mu} {\mkern 1mu} \,\,{\rm{for}}\,\,n = L/2,{\mkern 1mu} {\mkern 1mu}  \ldots L - 1.}
\end{array}} \right.{\mkern 1mu} }\\
{}\\
\begin{array}{l}
{\rm{When}}{\mkern 1mu} \,k'{\mkern 1mu} \,{\rm{is}}{\mkern 1mu} {\mkern 1mu} {\rm{odd}}:\\
Y_{k',n}^{\rm{De}} = 
\end{array}\\
\begin{array}{l}
\left\{ {\begin{array}{*{20}{l}}
{{\rm{0}},\,\,{\rm{for}}\,\,n = 0,{\mkern 1mu} {\mkern 1mu}  \ldots L/2 - 1}\\
{{v}_{0,n}^{k'}{\mkern 1mu} {D_{k',n}},{\mkern 1mu} \,\,{\rm{for}}\,\,n = L/2,{\mkern 1mu} {\mkern 1mu}  \ldots L - 1.}
\end{array}} \right.\\
\\
Y_{k',n}^{\rm{In}} = {\mkern 1mu} {\mkern 1mu} {\mkern 1mu} {\mkern 1mu} {\mkern 1mu} 
\end{array}\\
{{\mkern 1mu} \left\{ {\begin{array}{*{20}{l}}
{{v}_{ + 1,n}^{k'}{\mkern 1mu} {D_{k' - 1,n}} + {v}_{ - 1,n}^{k'}{\mkern 1mu} {D_{k' + 1,n}},{\mkern 1mu} {\mkern 1mu} \,{\rm{for}}\,\,n = 0,{\mkern 1mu} {\mkern 1mu}  \ldots L/2 - 1}\\
{0{\mkern 1mu} ,{\mkern 1mu} {\mkern 1mu} \,\,{\rm{for}}\,\,n = L/2,{\mkern 1mu} {\mkern 1mu}  \ldots L - 1.}
\end{array}} \right.}
\end{array}
\label{14}
\end{equation}

The structure of the proposed FBMC/OULP system for the ideal channel is shown in Fig. 4. To compensate the factor $v_{0,n}^{{{k}'}}$ (coefficients which associate with the desired part) in \eqref{14}, the pre-equalizer matrix ${{\mathbf{\Lambda }}_{k}}$ is used at the transmitter such that ${{\bar{D}}_{k}}\triangleq {{\mathbf{\Lambda }}_{k}}{{\bar{q}}_{k}}$. Where ${{\bar{q}}_{{{k}'}}}\triangleq {{[{{q}_{{k}',0}},{{q}_{{k}',1}},\ldots {{q}_{{k}',L/2-1}}]}^{\,T}}$ are the new version of transmitted symbols and ${{\mathbf{\Lambda }}_{k}}$ is a $L\times L/2$ matrix whose elements depend on the $k$ and are obtained from \eqref{12}. It is defined as
\begin{equation}
{{\bf{\Lambda }}_k} \buildrel \Delta \over = \left\{ \begin{array}{l}
{\left[ {{{\bf{E}}_k} \vdots {{\bf{0}}_{L/2}}} \right]^T}\,\,{\rm{ ,}}\,{\rm{\,\,for \,\, even \,\, }}k\\
{\left[ {{{\bf{0}}_{L/2}} \vdots {{\bf{E}}_k}} \right]^T}{\rm{ }}\,{\rm{,}}\,{\rm{\,\, for \,\, odd \,\,}}k,
\end{array} \right.
\label{15}
\end{equation}
where ${{\mathbf{E}}_{k}}\triangleq \text{diag}\left( 1/v_{0,0}^{k},\ldots 1/v_{0,L/2-1}^{k} \right)$ for even $k$ and  ${{\mathbf{E}}_{k}}\triangleq \text{diag}\left( 1/v_{0,L/2}^{k},\ldots 1/v_{0,L-1}^{k} \right)$ for odd $k$. In this equations, ${{\mathbf{0}}_{L/2}}$  indicated the zero matrix with size $L/2\times L/2$. Note that the pre-equalizer matrix ${{\mathbf{\Lambda }}_{k}}$ is designed based on the prototype filter features and does not depends on the channel information, and eventually; the transmitter does not need to know the channel. As it is seen in Fig. 4, after multiplication by ${{\mathbf{\Lambda }}_{k}}$, the complex-valued symbols ${{q}_{k,n}}$ are organized to be loaded on the lower or upper half of ${{\bar{D}}_{k}}$ in order to preserve the OULP transmission strategy shown in Fig. 2. Thus, with the pre-equalizer, \eqref{13} can be rewritten as
\begin{equation}
\begin{array}{*{20}{l}}
{{{\bar Y}_{k'}} = {\bf{V}}_0^{k'}{{\bf{\Lambda }}_{k'}}{\mkern 1mu} {{\bar q}_{k'}} + \sum\limits_{\begin{array}{*{20}{c}}
{\kappa  =  - 1}\\
{\kappa  \ne 0}
\end{array}}^1 {{\bf{V}}_\kappa ^{k'}{{\bf{\Lambda }}_{k' - \kappa }}{\mkern 1mu} {{\bar q}_{k' - \kappa }}{\mkern 1mu} }  + {{\bar \Omega }_{k'}}}\\
{{\mkern 1mu} {\mkern 1mu} {\mkern 1mu} {\mkern 1mu} {\mkern 1mu} {\mkern 1mu} {\mkern 1mu} {\mkern 1mu} {\mkern 1mu} {\mkern 1mu} \,\,\,\,\, = \bar Y_{k'}^{\rm{De}} + \bar Y_{k'}^{\rm{In}} + {{\bar \Omega }_{k'}},}
\end{array}
\label{16}
\end{equation}
As a result, the desired ($\bar{Y}_{{{k}'}}^{\rm{De}}$) and the interfered ($\bar{Y}_{{{k}'}}^{\rm{In}}$) parts are
\begin{equation}
\begin{array}{l}
{\rm{When}}\,\,k'\,\,{\rm{is}}\,\,{\rm{even:}}\\
Y_{k',n}^{\rm{De}} = \\
\left\{ \begin{array}{l}
{q_{k',n}}{\rm{ , for \,\, }}n = 0,\,\, \ldots L/2 - 1\\
0{\rm{ }},\,{\rm{for \,\, }}n = L/2,\,\, \ldots L - 1.
\end{array} \right.\\
\\
Y_{k',n}^{\rm{In}} = \\
\,\left\{ \begin{array}{l}
{\rm{0, for \,\, }}n = 0,\,\, \ldots L/2 - 1\\
\frac{{{v}_{ + 1,n}^{k'}\,}}{{{v}_{0,n}^{k' - 1}}}{q_{k' - 1,n - L/2}} + \frac{{{v}_{ - 1,n}^{k'}}}{{{v}_{0,n}^{k' + 1}}}\,{q_{k' + 1,n - L/2}}\,,\,\,\\
\,\,\,\,\,\,\,\,\,\,\,\,\,\,\,\,\,\,\,\,\,\,\,\,\,\,\,\,\,\,\,\,\,\,\,\,\,\,{\rm{for \,\, }}n = L/2,\,\, \ldots L - 1.
\end{array} \right.\\
\\
{\rm{When}}\,\,k'\,\,{\rm{is}}\,\,{\rm{odd:}}\\
Y_{k',n}^{\rm{De}} = \\
\left\{ \begin{array}{l}
{\rm{0}}\,{\rm{, for \,\,}}n = 0,\,\, \ldots L/2 - 1\\
{q_{k',n - L/2}},\,\,{\rm{for \,\, }}n = L/2,\,\, \ldots L - 1.
\end{array} \right.\,\\
\\
Y_{k',n}^{\rm{In}} = \,\\
\left\{ \begin{array}{l}
\frac{{{v}_{ + 1,n}^{k'}\,}}{{{v}_{0,n}^{k' - 1}}}{q_{k' - 1,n}} + \frac{{{v}_{ - 1,n}^{k'}}}{{{v}_{0,n}^{k' + 1}}}\,{q_{k' + 1,n}}\,{\rm{,}}\\
\,\,\,\,\,\,\,\,\,\,\,\,\,\,\,\,\,\,\,\,\,\,\,\,{\rm{for \,\, }}n = 0,\,\, \ldots L/2 - 1\\
0\,,\,{\rm{for \,\, }}n = L/2,\,\, \ldots L - 1.
\end{array} \right.
\end{array}
\label{17}
\end{equation}

\begin{figure*}[!t]
\centering
\includegraphics [width=6.5in]{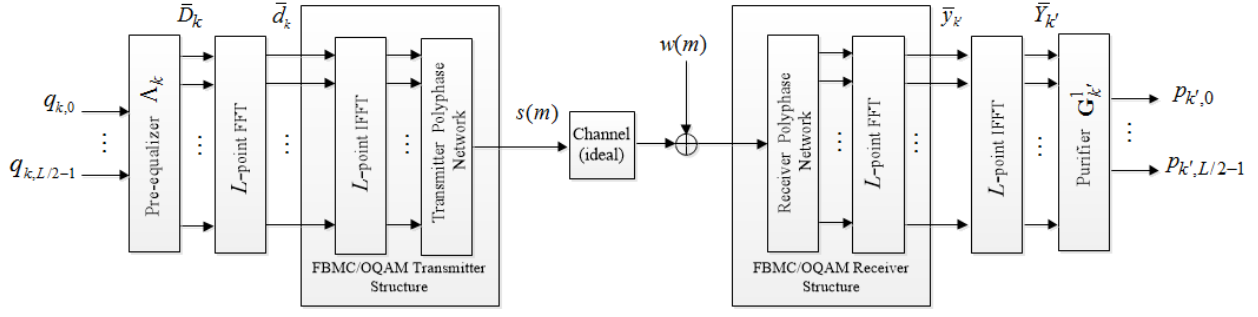}
\caption{The structure of the FBMC/OULP system when channel is flat.} 
\vspace{-3mm}
\label{F4}
\end{figure*}

As illustrated in Fig. 4, because of the isolation between desired and interfered parts in ${{\bar{Y}}_{{{k}'}}}$, to remove the interfered part, $\bar{Y}_{{{k}'}}^{\rm{In}}$ can be simply force to become zero by multiplying the purifier matrix $\mathbf{G}_{{{k}'}}^{1}$ as 
\begin{equation}
{{\bar{p}}_{{{k}'}}}\triangleq \mathbf{G}_{{{k}'}}^{1}{{\bar{Y}}_{{{k}'}}},
\label{18}
\end{equation}
in which ${{\bar{p}}_{{{k}'}}}$ is the purified version of ${{\bar{Y}}_{{{k}'}}}$ and $\mathbf{G}_{{{k}'}}^{1}$ is defined as
\begin{equation}
\begin{array}{l}
{\bf{G}}_{k'}^1 \buildrel \Delta \over = \left[ {\begin{array}{*{20}{c}}
{{{\bf{I}}_{L/2}}}&{{{\bf{0}}_{L/2}}}
\end{array}} \right],{\mkern 1mu} {\mkern 1mu} {\mkern 1mu} {\mkern 1mu} {\rm{for\,\,}}{\mkern 1mu} {\rm{even\,\,}}{\mkern 1mu} k'\\
{\bf{G}}_{k'}^1 \buildrel \Delta \over = \left[ {\begin{array}{*{20}{c}}
{{{\bf{0}}_{L/2}}}&{{{\bf{I}}_{L/2}}}
\end{array}} \right],{\mkern 1mu} {\mkern 1mu} {\mkern 1mu} {\mkern 1mu} {\rm{for\,\,}}{\mkern 1mu} {\rm{odd\,\,}}{\mkern 1mu} k',
\end{array}
\label{19}
\end{equation}
where ${{\mathbf{I}}_{L/2}}$ is the identity matrix with size $L/2\times L/2$. By considering \eqref{16} and \eqref{17} into \eqref{18}, ${{\bar{p}}_{{{k}'}}}$ becomes
\begin{equation}
{{\bar{p}}_{{{k}'}}}={{\bar{q}}_{{{k}'}}}+{{\bar{\psi }}_{{{k}'}}},
\label{20}
\end{equation}
where ${{\bar{p}}_{{{k}'}}}\triangleq {{[{{p}_{{k}',0}},{{p}_{{k}',1}},\ldots {{p}_{{k}',L/2-1}}]}^{\,T}}$ and the contaminating noise vector is defined as ${{\bar{\psi }}_{{{k}'}}}\triangleq \mathbf{G}_{{{k}'}}^{1}{{\bar{\Omega }}_{{{k}'}}}$. According to \eqref{20}, for every individual symbol we have
\begin{equation}
{{p}_{{k}',n}}={{q}_{{k}',n}}+{{\psi }_{{k}',n}}\,\,,\,\,\,\,\text{for}\,\,n=0,1,\ldots L/2-1.
\label{21}
\end{equation}
As \eqref{21} reveals, by using the proposed FBMC/OULP structure given in Fig. 4, the intrinsic interferences of the FBMC can be removed from the desired symbols in the final outcomes when the channel is flat.

\subsection{In frequency selective channels}
When the channel is not flat, the frequency selectivity of channel leads to an overlap between desired and interfered parts of the proposed system’s received signal. In this section we first elaborate the effect of the channel frequency selectivity on the FBMC/OULP outputs and then propose a method to eradicate it.

\subsubsection{The effect  of frequency selective channels}

Let us assume that in the $k\text{th}$ time slot, ${{h}_{k,m}}$ (for $m=0,1,\ldots {{L}_{\text{c}}}-1$) presents the CIR of the frequency selective channel with length ${{L}_{\text{c}}}$. Similar to FBMC/OQAM scheme, when the space between adjacent subchannels is smaller than the coherence bandwidth of the channel (which is guaranteed by a large value of $L$), by following \eqref{9}, the channel tainted output of matched-filters, ${{\bar{r}}_{{{k}'}}}$ can be given as \cite{R29}
\begin{equation}
{{\bar{r}}_{{{k}'}}}={{\mathbf{H}}_{{{k}'}}}\sum\limits_{\kappa=-1}^{1}{\mathbf{Z}_{\kappa}^{{{k}'}}\,}{{\bar{d}}_{{k}'-\kappa}}+{{\bar{\omega }}_{{{k}'}}},
\label{22}
\end{equation}
where $\mathbf{H}_{{{k}'}}^{{}}$ presents a diagonal matrix of CIR's FFT such that 
\begin{equation}
{{\mathbf{H}}_{{{k}'}}}=\text{diag}([{{H}_{{k}',0}},{{H}_{{k}',1}},\ldots {{H}_{{k}',L-1}}]),
\label{23}
\end{equation}
in which ${{[{{H}_{{k}',0}},{{H}_{{k}',1}},\ldots {{H}_{{k}',L-1}}]}^{T}}\triangleq {{\mathbf{F}}_{L}}{{\bar{h}}_{{{k}'}}}$ is the FFT of channel vector ${{\bar{h}}_{{{k}'}}}\triangleq [{{h}_{{k}',0}},\ldots {{h}_{{k}',{{L}_{\text{C}}}-1}},$ $0,\ldots 0{{]}^{T}}$ with size $L\times 1$. Following the procedure carried out to obtain \eqref{10}, after taking IFFT from both sides of \eqref{22}, it becomes
\begin{equation}
{{\bar{R}}_{{{k}'}}}={{\mathbf{U}}_{{{k}'}}}\,\sum\limits_{\kappa=-1}^{1}{\mathbf{V}_{\kappa}^{{{k}'}}{{{\bar{D}}}_{{k}'-\kappa}}\,}+{{\bar{\Omega }}_{{{k}'}}}.
\label{24}
\end{equation}
where ${{\bar{R}}_{{{k}'}}}\triangleq \mathbf{F}_{L}^{\dagger }{{\bar{r}}_{{{k}'}}}$ and ${{\mathbf{U}}_{{{k}'}}}\triangleq \mathbf{F}_{L}^{\dagger }{{\mathbf{H}}_{{{k}'}}}\,{{\mathbf{F}}_{L}}$ is a circular $L\times L$ matrix of ${{h}_{{k}',m}}$ coefficients, for $m=0,1,\ldots {{L}_{\text{c}}}-1$ (see Appendix B)
\begin{equation}
{{\bf{U}}_{k'}} = \left[ {\begin{array}{*{20}{c}}
{{h_{k',0}}}&{\bf{0}}&{{h_{k',{L_{\rm{c}}} - 1}}}& \cdots &{{h_{k',1}}}\\
 \vdots & \searrow &{\bf{0}}& \searrow & \vdots \\
{{h_{k',{L_{\rm{c}}} - 1}}}&{}&{{h_{k',0}}}&{}&{{h_{k',{L_{\rm{c}}} - 1}}}\\
{}& \searrow & \vdots & \searrow &{\bf{0}}\\
{\bf{0}}&{}&{{h_{k',{L_{\rm{c}}} - 1}}}& \cdots &{{h_{k',0}}}
\end{array}} \right],
\label{25}
\end{equation}
According to \eqref{16}, we can represent \eqref{24} with $\bar{Y}_{{{k}'}}^{\rm{De}}$ and $\bar{Y}_{{{k}'}}^{\rm{In}}$, such that
\begin{equation}
{{\bar{R}}_{{{k}'}}}={{\mathbf{U}}_{{{k}'}}}\,(\bar{Y}_{{{k}'}}^{\rm{De}}+\bar{Y}_{{{k}'}}^{\rm{In}})+{{\bar{\Omega }}_{{{k}'}}}.
\label{26}
\end{equation}

As it can be seen, in contrast to \eqref{16}, when channel is not flat, multiplication of ${{\mathbf{U}}_{{{k}'}}}\,$ causes a circular overlap among $\bar{Y}_{{{k}'}}^{\rm{De}}$ and $\bar{Y}_{{{k}'}}^{\rm{In}}$ at each time slot ${k}'$ (see \eqref{17} and the lattice structure in Fig. 3). Considering the structure of $\,{{\mathbf{U}}_{{{k}'}}}$ in \eqref{25}, it is clear that the leakage of non-zero parts of $\bar{Y}_{{{k}'}}^{\rm{De}}$ and $\bar{Y}_{{{k}'}}^{\rm{In}}$, depends on ${{L}_{\text{c}}}$. Total amount of this leakage is $2{{L}_{\text{c}}}-2$, including ${{L}_{\text{c}}}-1$ overlap of $Y_{{k}',n}^{\rm{In}}$ through $Y_{{k}',n}^{\rm{De}}$ from one side and ${{L}_{\text{c}}}-1$ overlap of $Y_{{k}',n}^{\rm{De}}$ through $Y_{{k}',n}^{\rm{In}}$ from the other side. One way to get rid of this disturbance is to append a CP to transmitted symbols in each time slot. However, this CP length should be at least ${{L}_{\text{c}}}-1$ which diminishes the bandwidth efficiency of the FBMC. However, because of the circular structure of ${{\mathbf{U}}_{{{k}'}}}$, there is a potential to eliminate the overlap between desired and interfered parts without considering any CP. In this regard, we propose the interference eliminator (IE), to estimate and mitigate the interfered symbols that overlap with the desired part due to the frequency selectivity of channel. 

\subsubsection{The IE structure}

Since the channel's FFT appears as a diagonal matrix ${{\mathbf{H}}_{{{k}'}}}$ in \eqref{22}, it can be removed by utilizing an MMSE estimator. Thus, according to the IE structure, presented in Fig. 5, the output of the MMSE estimator yields to
\begin{equation}
{{\hat{\bar{y}}}_{{{k}'}}}=\mathbf{H}_{{{k}'}}^{+}\,{{\bar{r}}_{{{k}'}}}
\label{27}
\end{equation}
where the MMSE matrix is defined as $\mathbf{H}_{{{k}'}}^{+}\triangleq {{[\mathbf{H}_{{{k}'}}^{\dagger }{{\mathbf{H}}_{{{k}'}}}+{{N}_{0}}{{\mathbf{I}}_{L}}]}^{\,-1}}\,\mathbf{H}_{{{k}'}}^{\dagger }$, in which ${{(.)}^{\dagger }}$ indicates the Hermitian operator. As is obvious ${{\hat{\bar{y}}}_{{{k}'}}}$ is an estimation of ${{\bar{y}}_{{{k}'}}}$ in \eqref{9}. After taking an $L$-point IFFT of ${{\hat{\bar{y}}}_{{{k}'}}}$, it results in an estimation of ${{\bar{Y}}_{{{k}'}}}$ in \eqref{10}, which we denote it by  ${{\hat{\bar{Y}}}_{{{k}'}}}\triangleq \mathbf{F}_{L}^{\dagger }{{\hat{\bar{y}}}_{{{k}'}}}=\mathbf{F}_{L}^{\dagger }\mathbf{H}_{{{k}'}}^{+}\,{{\bar{r}}_{{{k}'}}}$. Similar to ${{\bar{Y}}_{{{k}'}}}$, there is no overlap between desired ($\hat{\bar{Y}}_{{k}',m}^{\rm{De}}$) and interfered ($\hat{\bar{Y}}_{{k}',m}^{\rm{In}}$) parts of ${{\hat{\bar{Y}}}_{{{k}'}}}$. Consequently, the interfered part can be separated as
\begin{equation}
\hat{\bar{Y}}_{{{k}'}}^{\rm{In}}\triangleq {{\mathbf{\Pi }}_{{{k}'}}}\,{{\hat{\bar{Y}}}_{{{k}'}}}={{\mathbf{\Pi }}_{{{k}'}}}\mathbf{F}_{L}^{\dagger }\mathbf{H}_{{{k}'}}^{+}\,{{\bar{r}}_{{{k}'}}},
\label{28}
\end{equation}
where the selector matrix ${{\mathbf{\Pi }}_{{{k}'}}}$, for even and odd quantities of ${k}'$, is defined as
\begin{equation}
\begin{array}{l}
{{\bf{\Pi }}_{k'}} \buildrel \Delta \over = \left[ {\begin{array}{*{20}{c}}
{{{\bf{0}}_{L/2}}}&{{{\bf{0}}_{L/2}}}\\
{{{\bf{0}}_{L/2}}}&{{{\bf{I}}_{L/2}}}
\end{array}} \right],{\mkern 1mu} {\mkern 1mu} {\mkern 1mu} {\mkern 1mu} {\rm{for \,\,}}{\mkern 1mu} {\rm{even \,\,}}{\mkern 1mu} k';{\mkern 1mu} {\mkern 1mu} \\
\\
{\mkern 1mu} {{\bf{\Pi }}_{k'}} \buildrel \Delta \over = \left[ {\begin{array}{*{20}{c}}
{{{\bf{I}}_{L/2}}}&{{{\bf{0}}_{L/2}}}\\
{{{\bf{0}}_{L/2}}}&{{{\bf{0}}_{L/2}}}
\end{array}} \right],{\mkern 1mu} {\mkern 1mu} {\mkern 1mu} {\mkern 1mu} {\rm{for\,\,}}{\mkern 1mu} {\rm{odd\,\,}}{\mkern 1mu} k'.
\end{array}
\label{29}
\end{equation}

After multiplication of the FFT of the interference vector $\hat{\bar{Y}}_{{{k}'}}^{\rm{In}}$ by ${{\mathbf{H}}_{{{k}'}}}$ and subtracting it from ${{\bar{r}}_{{{k}'}}}$, as shown in Fig. 5, the output of the IE block, ${{\tilde{\bar{r}}}_{{{k}'}}}$, can be expressed as
\begin{equation}
{{\tilde{\bar{r}}}_{{{k}'}}}\triangleq {{\bar{r}}_{{{k}'}}}-{{\mathbf{H}}_{{{k}'}}}\,{{\mathbf{F}}_{L}}\,\hat{\bar{Y}}_{{{k}'}}^{\rm{In}}.
\label{30}
\end{equation}
Hence ${{\tilde{\bar{r}}}_{{{k}'}}}$ is the reduced interference version of ${{\bar{r}}_{{{k}'}}}$ and it can be used to detect the desired symbols transmitted over frequency selective channel.

\begin{figure*}[!t]
\centering
\includegraphics [width= 4.3 in]{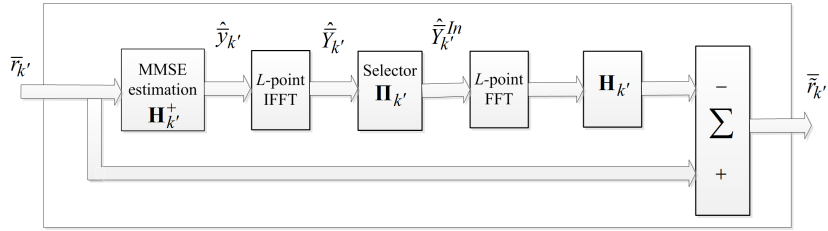}
\caption{The structure of IE.} 
\vspace{-3mm}
\label{F5}
\end{figure*}

\subsubsection{The OULP/FBMC system for frequency selective channels}

Fig. 6 shows the structure of the FBMC/OULP system to extract the desired symbols when channel is frequency selective (non-flat). As it can be seen, the IE block is appended at the receiver. Note that, as it is shown in Fig. 4, the polyphase networks cascaded to the $L$-point IFFT and $L$-point FFT can be used to implement the FBMC/OQAM transmitter and receiver structures, respectively \cite{R8}. Thus in Fig. 4, the FFT and the IFFT blocks can be canceled at the transmitter side of the FBMB/OULP system and only the polyphase network remains. However, in Fig. 6 due to using the IE structure between the FFT and IFFT blocks at the receiver side, we cannot simplify the receiver structure of the FBMC/OULP system. It is noteworthy that the appellation of offset upper-lower polyphase network (OULP) system for the proposed scheme, presented in Fig. 6, is originated form this fact that at the transmitter of the proposed system the symbols are alternatively loaded on the upper and lower halves of the polyphase network branches.

Nevertheless, according to \eqref{30}, after taking an IFFT from the IE output, it yields to 
\begin{equation}
\begin{array}{l}
{\tilde{ \bar R}}_{k'} = {{\bf{F}}^\dag }({{\bar r}_{k'}} - {{\bf{H}}_{k'}}\,{{\bf{F}}_L}\,{\hat {\bar Y}}_{k'}^{\rm{In}})\\
\,\,\,\,\,\,\,\,\,\,\, = {{\bar R}_{k'}} - {{\bf{U}}_{k'}}{\hat {\bar Y}}_{k'}^{\rm{In}}.
\end{array}
\label{31}
\end{equation}
Substituting \eqref{26} into \eqref{31} and assuming a perfect estimation of interference ($\hat{\bar{Y}}_{{{k}'}}^{\rm{In}}=\bar{Y}_{{{k}'}}^{\rm{In}}$), it yields to
\begin{equation}
{{\tilde{\bar{R}}}_{{{k}'}}}={{\mathbf{U}}_{{{k}'}}}\bar{Y}_{{{k}'}}^{\rm{De}}+{{\bar{\Omega }}_{{{k}'}}}.
\label{32}
\end{equation}
Since, from \eqref{17}, for even (or odd) ${k}'$ the upper (or lower) half of $\bar{Y}_{{{k}'}}^{\rm{De}}$ is zero, it is expected that the $n\text{th}$ element of vector ${{\tilde{\bar{R}}}_{{{k}'}}}$ for $n=L/2+{{L}_{\text{c}}}-1,\,\,\ldots \,L-1$ (or for $n={{L}_{\text{c}}}-1,\ldots L/2-1$) become zero (refer to the structure of ${{\mathbf{U}}_{{{k}'}}}$ in \eqref{25}). However, because of the noise existence and imperfect interference mitigation by the IE, those elements may not be precisely zeroes. Thus, according to Fig. 6, they can be forced to become zero, by using the purifier matrix $\mathbf{G}_{{{k}'}}^{{{L}_{\text{c}}}}$. In this way, we have
\begin{equation}
{{\bar{x}}_{{{k}'}}}=\mathbf{G}_{{{k}'}}^{{{L}_{\text{c}}}}\,{{\tilde{\bar{R}}}_{{{k}'}}}=\mathbf{G}_{{{k}'}}^{{{L}_{\text{c}}}}\,{{\mathbf{U}}_{{{k}'}}}\bar{Y}_{{{k}'}}^{\rm{De}}+{{\bar{\psi }}_{{{k}'}}},
\label{33}
\end{equation}
where ${{\bar{\psi }}_{{{k}'}}}\triangleq \mathbf{G}_{{{k}'}}^{{{L}_{\text{c}}}}\,{{\bar{\Omega }}_{{{k}'}}}$ is the output noise and $\mathbf{G}_{{{k}'}}^{{{L}_{\text{c}}}}$ is defined as
\begin{equation}
\begin{array}{l}
{\bf{G}}_{k'}^{{L_{\rm{c}}}} \buildrel \Delta \over = \\
\,\,\,\,\,\,\,\left[ {\begin{array}{*{20}{c}}
{{{\bf{I}}_{L/2}}}&{\begin{array}{*{20}{c}}
{{{\bf{I}}_{{L_{\rm{c}}} - 1}}}&{}\\
{{{\bf{0}}_{(L/2 - {L_{\rm{c}}} + 1) \times ({L_{\rm{c}}} - 1)}}}&{}
\end{array}{{\bf{0}}_{(L/2) \times (L/2 - {L_{\rm{c}}} + 1)}}}
\end{array}} \right],\,\\
\,\,\,\,\,\,\,{\rm{for}}\,{\rm{even}}\,k';\\
\\
{\bf{G}}_{k'}^{{L_{\rm{c}}}} \buildrel \Delta \over = \\
\,\,\,\,\,\,\,\,\left[ {\begin{array}{*{20}{c}}
{\begin{array}{*{20}{c}}
{{{\bf{I}}_{{L_{\rm{c}}} - 1}}}&{}\\
{{{\bf{0}}_{(L/2 - {L_{\rm{c}}} + 1) \times ({L_{\rm{c}}} - 1)}}}&{}
\end{array}{{\bf{0}}_{(L/2) \times (L/2 - {L_{\rm{c}}} + 1)}}}&{{{\bf{I}}_{L/2}}}
\end{array}} \right],\,\,\,\\
\,\,\,\,\,\,\,\,{\rm{for}}\,{\rm{odd}}\,k'.
\end{array}
\label{34}
\end{equation}

Note that the purifier matrix $\mathbf{G}_{{{k}'}}^{{{L}_{\text{c}}}}$ extremely reduces the residual interfered part and noise effects after performing the MMSE estimation in the IE block. To make it more clear, let’s assume that the MMSE estimation of $\hat{\bar{Y}}_{{{k}'}}^{\rm{In}}$ is imperfect and in \eqref{31} $\hat{\bar{Y}}_{{{k}'}}^{\rm{In}}$ is not exactly equal to $\bar{Y}_{{{k}'}}^{\rm{In}}$. This difference, which is the residual interfered part, is defined as ${{\bar{\Gamma }}_{{{k}'}}}\triangleq \bar{Y}_{{{k}'}}^{\rm{In}}-\hat{\bar{Y}}_{{{k}'}}^{\rm{In}}$ (note that the lower or upper half of ${{\bar{\Gamma }}_{{{k}'}}}$ is zero if ${k}'$ is even or odd, respectively) . As a result, instead of \eqref{32}, we have ${{\tilde{\bar{R}}}_{{{k}'}}}={{\mathbf{U}}_{{{k}'}}}(\bar{Y}_{{{k}'}}^{\rm{De}}+{{\bar{\Gamma }}_{{{k}'}}})+{{\bar{\Omega }}_{{{k}'}}}$. After the purification in \eqref{33}, it yields to ${{\bar{x}}_{{{k}'}}}=\mathbf{G}_{{{k}'}}^{{{L}_{\text{c}}}}\,{{\mathbf{U}}_{{{k}'}}}\bar{Y}_{{{k}'}}^{\rm{De}}+$ $\mathbf{G}_{{{k}'}}^{{{L}_{\text{c}}}}{{\mathbf{U}}_{{{k}'}}}\,{{\bar{\Gamma }}_{{{k}'}}}+{{\bar{\psi }}_{{{k}'}}}$. It is obvious that in term $\mathbf{G}_{{{k}'}}^{{{L}_{\text{c}}}}{{\mathbf{U}}_{{{k}'}}}\,{{\bar{\Gamma }}_{{{k}'}}}$ there are just ${{L}_{\text{c}}}-1$ non-zero elements and affect the detection of the next steps. On the other hand, since ${{L}_{\text{c}}}\ll \,\,L$, it can be concluded that, after purification, the residual interference impact on ${{\bar{x}}_{{{k}'}}}$ is extremely trivial. Fig. 7 shows the purification procedure for even ${k}'$. 

Following the procedure of Fig. 6, by substituting \eqref{17} into \eqref{33}, ${{\bar{x}}_{{{k}'}}}$ can be presented as the multiplication of ${{\mathbf{u}}_{{{k}'}}}$ by the vector of transmitted symbols 
\begin{equation}
{{\bar{x}}_{{{k}'}}}={{\mathbf{u}}_{{{k}'}}}{{\bar{q}}_{{{k}'}}}+{{\bar{\psi }}_{{{k}'}}},
\label{35}
\end{equation}
where, as it is shown in Appendix C, ${{\mathbf{u}}_{{{k}'}}}$ is a circular $(L/2)\times (L/2)$ matrix of ${{h}_{{k}',m}}$ coefficients 
\begin{equation}
{{\bf{u}}_{k'}} = \left[ {\begin{array}{*{20}{c}}
{{h_{k',0}}}&{\bf{0}}&{{h_{k',{L_{\rm{c}}} - 1}}}& \cdots &{{h_{k',1}}}\\
 \vdots & \searrow &{\bf{0}}& \searrow & \vdots \\
{{h_{k',{L_{\rm{c}}} - 1}}}&{}&{{h_{k',0}}}&{}&{{h_{k',{L_{\rm{c}}} - 1}}}\\
{}& \searrow & \vdots & \searrow &{\bf{0}}\\
{\bf{0}}&{}&{{h_{k',{L_{\rm{c}}} - 1}}}& \cdots &{{h_{k',0}}}
\end{array}} \right].
\label{36}
\end{equation}

According to the Fig. 6, after taking the $L/2\text{-point}$ FFT of both sides of \eqref{35}, the final outcome becomes
\begin{equation}
{{\bar{\chi }}_{{{k}'}}}=\,{{\mathbf{h}}_{{{k}'}}}\,{{\bar{Q}}_{{{k}'}}}+{{\bar{\Psi }}_{{{k}'}}}
\label{37}
\end{equation}
where ${{\bar{\chi }}_{{{k}'}}}={{\mathbf{F}}_{L/2}}{{\bar{x}}_{{{k}'}}}$, ${{\bar{Q}}_{{{k}'}}}={{\mathbf{F}}_{L/2}}{{\bar{q}}_{{{k}'}}}$ and ${{\bar{\Psi }}_{{{k}'}}}={{\mathbf{F}}_{L/2}}{{\bar{\psi }}_{{{k}'}}}$. Also ${{\mathbf{h}}_{{{k}'}}}\triangleq $ ${{\mathbf{F}}_{L/2}}{{\mathbf{u}}_{{{k}'}}}\mathbf{F}_{L/2}^{\dagger }$ which, as shown in Appendix D, is a diagonal matrix constructed by $L/2\text{-point}$ FFT elements of CIR coefficients 
\begin{equation}
{{\mathbf{h}}_{{{k}'}}}=\text{diag}([{{\mathcal{H}}_{{k}',0}},{{\mathcal{H}}_{{k}',1}},\ldots {{\mathcal{H}}_{{k}',L/2-1}}]),
\label{38}
\end{equation}
where ${{[{{\mathcal{H}}_{{k}',0}},{{\mathcal{H}}_{{k}',1}},\ldots {{\mathcal{H}}_{{k}',L/2-1}}]}^{T}}\triangleq {{\mathbf{F}}_{L/2}}{{\bar{h}}_{{{k}'}}}$ is the $L/2\text{-point}$ FFT of ${{\bar{h}}_{{{k}'}}}\triangleq [{{h}_{{k}',0}},\ldots {{h}_{{k}',{{L}_{\text{C}}}-1}},0,\ldots $ $0{{]}^{T}}$with size $L/2\times 1$. As \eqref{37} indicates, since ${{\bar{\chi }}_{{{k}'}}}={{[{{\chi }_{{k}',0}},{{\chi }_{{k}',1}},\ldots {{\chi }_{{k}',L/2-1}}]}^{T}}$, ${{\bar{Q}}_{{{k}'}}}=[{{Q}_{{k}',0}},$ ${{Q}_{{k}',1}},\ldots {{Q}_{{k}',L/2-1}}{{]}^{T}}$and ${{\bar{\Psi }}_{{{k}'}}}=[{{\Psi }_{{k}',0}},{{\Psi }_{{k}',1}}$, $\ldots {{\Psi }_{{k}',L/2-1}}{{]}^{T}}$, for every individual symbol $\eta \text{th}$, \eqref{37} yields to
\begin{equation}
{{\chi }_{{k}',\eta }}=\,{{\mathcal{H}}_{{k}',\eta }}\,{{Q}_{{k}',\eta }}+{{\Psi }_{{k}',\eta }}, \,\, \text{for}\,\,\eta =0,1,\ldots L/2-1.
\label{39}
\end{equation}

As is obvious, at the receiver of FBMC/OULP the received ${{\chi }_{{k}',\eta }}$ is the multiplication of the complex-value symbol ${{Q}_{{k}',\eta }}$ (the transmitted at the ${k}'\text{th}$ time slot over the $\eta \text{th}$ subchannel) by the $\eta \text{th}$ FFT component of CIR, contaminated by noise. Since $L/2\text{-point}$ FFT appeared in \eqref{37}, it can be interpreted that the complex-value symbols (${{Q}_{{k}',\eta }}$’s) are transmitted in time and frequency positions when $\Delta T=L/2$ and $\Delta F=2/L$. Thus, for this time-frequency lattice, the data symbol density becomes $\beta =1/(\Delta T\times \,\Delta F)=1$. According to \eqref{39}, in the proposed FBMC/OULP system when channel is frequency selective, ${{Q}_{{k}',\eta }}$ can be detected by using a simple ML detector after receiving ${{\chi }_{{k}',n}}$.

\begin{figure*}[!t]
\centering
\includegraphics [width=7in]{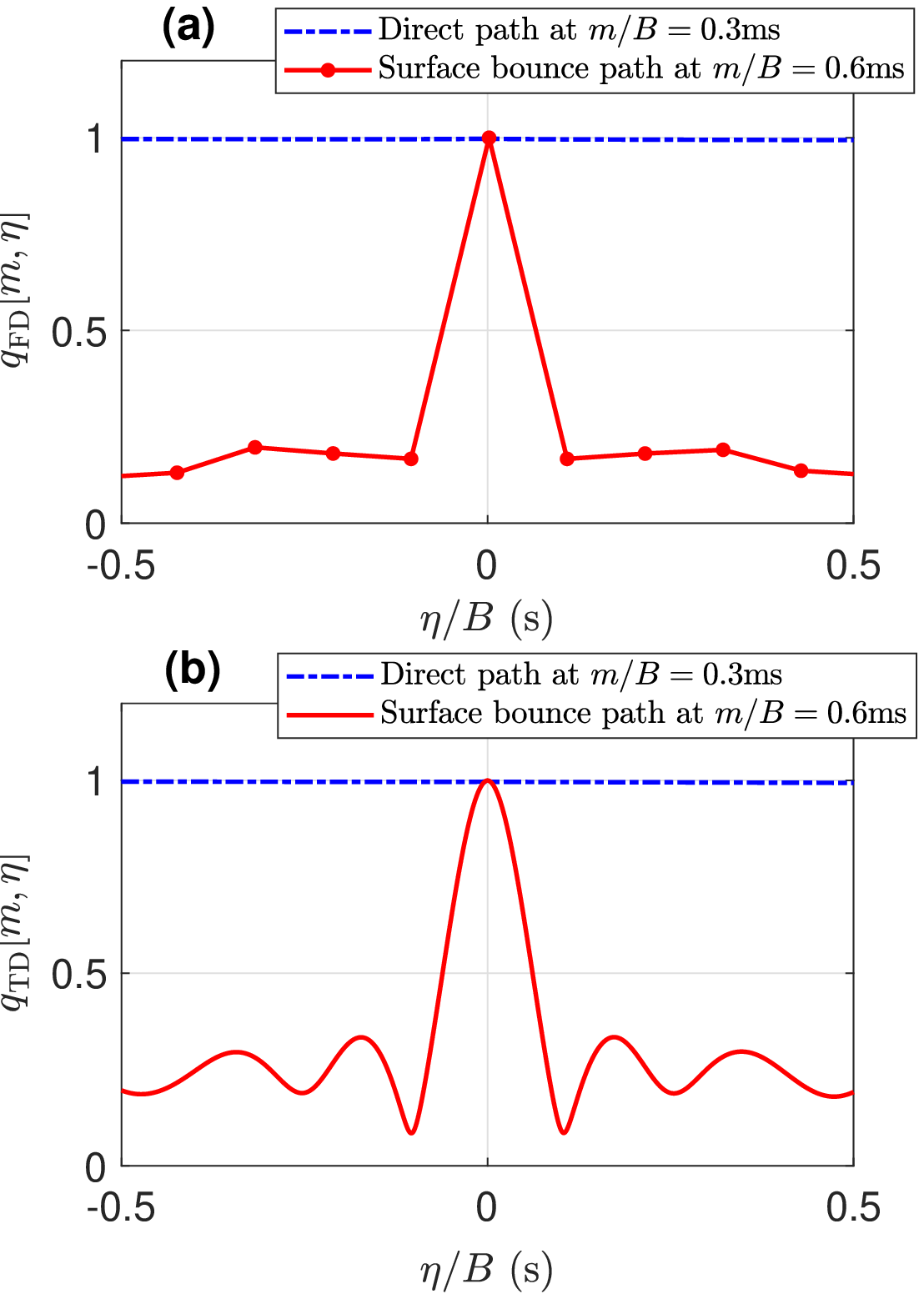}
\vspace{-0mm}
\caption{The structure of the FBMC/OULP system when channel is frequency selective.} \vspace{-0mm}
 \label{F6}
\end{figure*}

\begin{figure*}[!t]
\centering
\includegraphics [width=4in]{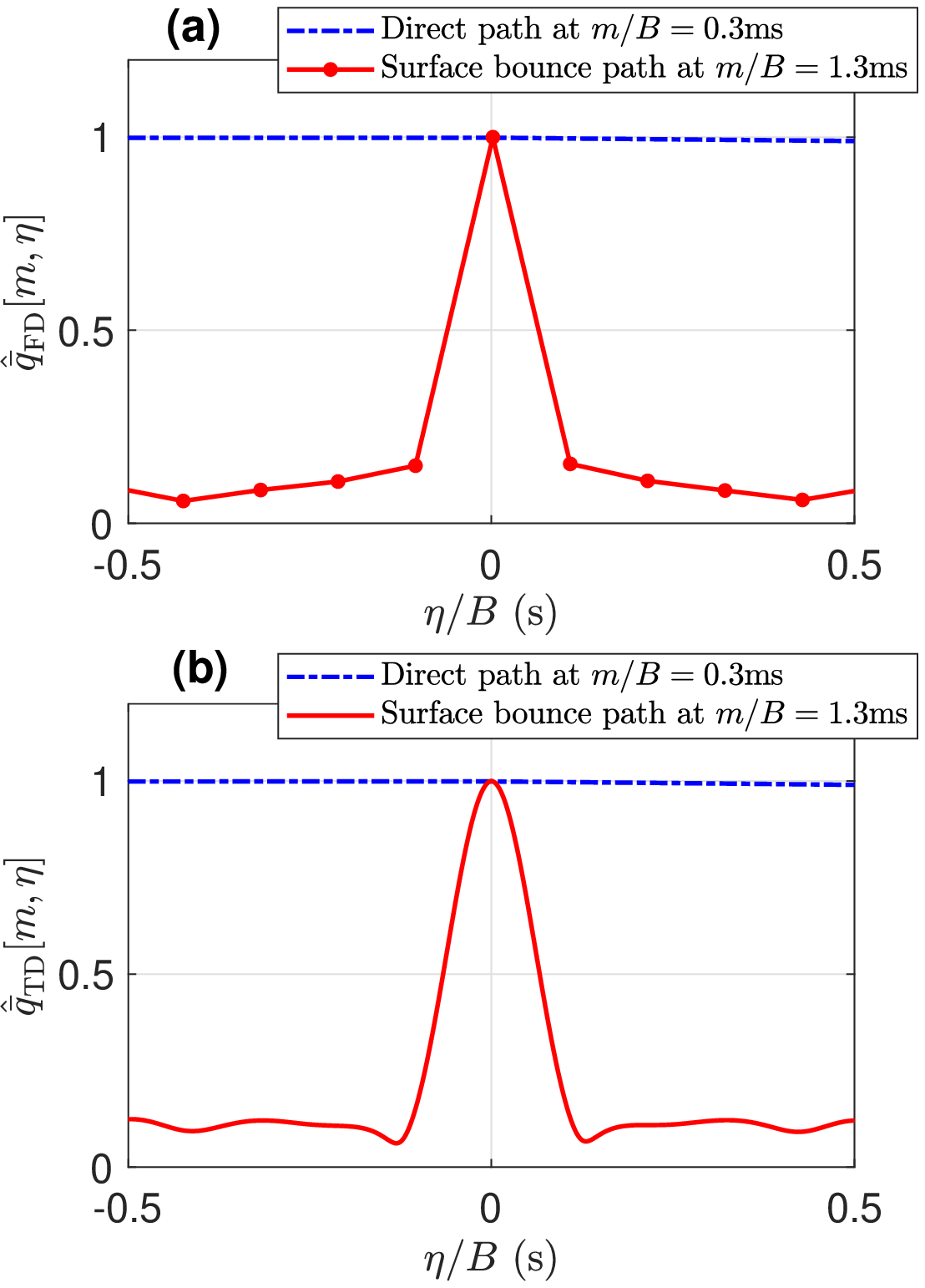}
\vspace{-0mm}
\caption{(a) For an even ${k}'$, the procedure of purification over ${{\bar{\tilde{R}}}_{{{k}'}}}$ to remove the residual interferences and noise and also preserve the circularity; (b) the resulted vector after purification ${{\bar{x}}_{{{k}'}}}$ with size $L/2\times 1$.} 
\vspace{-3mm}
 \label{F7}
\end{figure*}

\section{THE FBMC/OULP SYSTEM FOR MIMO CHANNELS}
In this section, we extend the proposed FBMC/OULP scheme for MIMO system with ${{N}_{\text{t}}}$ transmitter and ${{N}_{\text{r}}}$ receiver antennas. At the ${k}'\text{th}$ time slot and $\eta \text{th}$ subchannel, consider $\mathcal{H}_{\,{k}',\eta }^{ji}$, $Q_{{k}',\eta }^{i}$, $\chi _{{k}',\eta }^{j}$ and $\Psi _{{k}',\eta }^{j}$ as the channel $L/2$-point FFT coefficient between $i\text{th}$ transmitter antenna and $j\text{th}$ receiver antenna, the transmitted symbol by the $i\text{th}$ transmitter antenna, the output signal resulted from the $j\text{th}$ receiver antenna and the contaminating noise, respectively. According to \eqref{39}, the output matrix of the MIMO FBMC/OULP system becomes
\begin{equation}
\begin{array}{l}
\left[ {\begin{array}{*{20}{c}}
{\chi _{k',\eta }^1}& \ldots &{\chi _{k' + K,\eta }^1}\\
 \vdots & \ddots & \vdots \\
{\chi _{k',\eta }^{{N_{\rm{r}}}}}& \cdots &{\chi _{k' + K,\eta }^{{N_{\rm{r}}}}}
\end{array}} \right] = \\
\left[ {\begin{array}{*{20}{c}}
{{\cal H}_{\,k',\eta }^{11}}& \ldots &{{\cal H}_{\,k',\eta }^{1{N_{\rm{t}}}}}\\
 \vdots & \ddots & \vdots \\
{{\cal H}_{\,k',\eta }^{{N_{\rm{r}}}1}}& \ldots &{{\cal H}_{\,k',\eta }^{{N_{\rm{r}}}{N_{\rm{t}}}}}
\end{array}} \right].\left[ {\begin{array}{*{20}{c}}
{Q_{k',\eta }^1}& \ldots &{Q_{k' + K,\eta }^1}\\
 \vdots & \ddots & \vdots \\
{Q_{k',\eta }^{{N_{\rm{t}}}}}& \ldots &{Q_{k' + K,\eta }^{{N_{\rm{t}}}}}
\end{array}} \right]\\
\,\,\,\,\,\,\,\,\,\,\,\,\,\,\,\,\,\,\,\,\,\,\,\,\,\,\,\,\,\,\,\,\,\,\,\,\,\,\,\,\,\,\,\,\,\,\,\,\,\,\,\,\,\,\,\,\,\,\,\,\,\,\,\,\, + \left[ {\begin{array}{*{20}{c}}
{\Psi _{k',\eta }^1}& \ldots &{\Psi _{k' + K,\eta }^1}\\
 \vdots & \ddots & \vdots \\
{\Psi _{k',\eta }^{{N_{\rm{r}}}}}& \ldots &{\Psi _{k' + K,\eta }^{{N_{\rm{r}}}}}
\end{array}} \right],
\end{array}
\label{40}    
\end{equation}
where $K$ is the time duration of the transmission process in which the channel coefficients are assumed to be constant. For more facility, we can express  \eqref{40} as
\begin{equation}
{{\mathbf{\chi }}_{\eta }}={{\bf{\Xi }}_{\eta }}{{\mathbf{Q}}_{\eta }}+{{\mathbf{\Psi }}_{\eta }},
\label{41}    
\end{equation}
in which the time slot subscriptions are ignored. The ML detection of ${{\bar{Q}}_{\eta }}$ (the vector of all symbols included in ${{\mathbf{Q}}_{\eta }}$) is obtained by
\begin{equation}
{{\bar{Q}}_{\eta \,\,ML}}=\underset{\bar{Q}\in \,\varsigma }{\mathop{\arg \min }}\,\left\{ {{\left\| \,\,{{\mathbf{\chi }}_{\eta }}-{{\mathbf{\Xi}}_{\eta }}{{\mathbf{Q}}_{\eta }} \right\|}^{2}} \right\},
\label{42}
\end{equation}
where $\varsigma $ is the collection of all possible scenarios of ${{\bar{Q}}_{n}}$ and $\left\| \,.\, \right\|$ presents the Euclidean norm. If ${{\mathbf{Q}}_{\eta }}$ is organized as an OSTBC, the ML detection of \eqref{42} is simply performed for every single symbol separately. Thus, because of using complex-valued symbols along with employing the ML detection, the full diversity gain of MIMO channel can be achieved in the FBMC/OULP system. In addition, because of non-CP strategy, FBMC/OULP is more bandwidth efficient than the OFDM. 

Note that the ML detection method of \eqref{42} is based on the time domain independence of zero mean Gaussian noise elements of matrix ${{\mathbf{\Psi }}_{\eta }}$. Since the CIR length is generally less than the number of subchannels in the FBMC systems, in Appendix E, we show that the correlation between elements of ${{\mathbf{\Psi }}_{\eta }}$, regarding the time index, is less than 0.3 (for the case of the IOTA prototype filter). This quantity is considered as a weak correlation \cite{R34} and has an insignificant impact on practical scenes in the ML symbol detection. 

\section{Simulation Results}
To evaluate the performance of the proposed FBMC/OULP system, we consider this multicarrier scheme with different numbers of subchannels $L=128,\,\,256$ and $\text{512}$. Moreover, for comparisons, the OFDM and FFT-FBMC (proposed in \cite{R35}) multicarrier systems are simulated, as well. The multi-path channel has the following power delay profile (PDP) \\

$\begin{array}{l}
{\rm{PDP}} = [0\,\, - 1\,\,\, - 9\,\,\, - 10\,\,\, - 15\,\, - 20]\,\,{\rm{dB}};\\
{\rm{Delay}} = [0\,\,\,100\,\,\,300\,\,500\,\,\,800\,\,\,1300]\,\,{\rm{ns}}.
\end{array}$\\

\noindent This channel model is similar to the Vehicular-A model given in \eqref{40} with the modified delays. Table II presents other parameters of the simulations. As shown in this table, different channel coherence times ${{T}_{\text{co}}}=80\,\text{ms},\text{ }800\,\text{ }\!\!\mu\!\!\text{ s}$ and $80\,\text{ }\!\!\mu\!\!\text{ s}$ are considered. At the rest of the paper, we  mention the channel with these coherence times as the slow, fast and very fast multi-path (SM, FM, VFM) fading channels, respectively. Note that in this system, when $L=128,\,\,256$ and $\text{512}$, the time offset and frequency space are $\Delta T={{t}_{s}}L/2=6400,\,\,12800$ and $25600\,\text{ns}$ and $\Delta F= 1/(L{{t}_{s}})=78125,\,\,39062$ and $19531$ Hz, respectively.

\begin{table}
\caption{OFDM parameters}
\centering
\begin{tabular}{ |c|c|c|c|c|c|c|c|c|c|c|c|c|c| } 
\hline 
modulation & 16-QAM \\ 
\hline
prototype filter  & IOTA  \\ 
\hline
sampling period & ${{t}_{\text{s}}}=1\text{00}~\,\text{ns}~$\\ 
\hline
carrier frequency & ${{f}_{\text{c}}}=1\,\,\text{GHz}$  \\ 
\hline  
OFDM CP length & $\upsilon =12$ \\
\hline
FFT-FBMC per-subchannel-OFDM size & $N=16,\text{ }32,\text{ 64}$ \\
\hline
FFT-FBMC CP length  & ${\upsilon }'=2$ \\
\hline
\# of subcarriers  & $L=128,\,\,256,\,\,512$\\ 
\hline 
OFDM symbol length  & $14,\,\,27,\,\,53\,\,\text{ }\!\!\mu\!\!\text{ s}$  \\ 
\hline
FFT-FBMC symbol length  & $172.8,\,\,345.6,\,\,691.2\,\,\,\text{ }\!\!\mu\!\!\text{ s}$  \\ 
\hline
FBMC/OULP symbol length  & $12.8,\,\,25.6,\,\,51.2\,\,\text{ }\!\!\mu\!\!\text{ s}$  \\
\hline
channel coefficients’ correlation time & ${{T}_{\text{co}}}=80\,\text{ms,}\,\,800\,\text{ }\!\!\mu\!\!\text{ s},\,\,80\,\text{ }\!\!\mu\!\!\text{ s}$ \\ 
\hline
number of MIMO antennas & ${{N}_{\text{t}}}={{N}_{\text{r}}}=2$ \\ 
\hline
type of STBC & Alamouti  \\ 
\hline
\end{tabular}
\label{T2}
\end{table} 
According to Table II, when $\upsilon $ denotes the length of the CP, $\mu \triangleq \upsilon /(\upsilon +L)\times 100\%$ is defined as the bandwidth efficiency loss. Under defined parameters in Table II, for the OFDM system $\upsilon =12$; thus ${{\mu }_{\text{OFDM}}}=8.6\%,\text{ 4}\text{.5 }\!\!\%\!\!\text{ }$ and $\text{2}\text{.3 }\!\!\%\!\!\text{ }$ when $L=128$, $\text{ }256$ and $\text{512}$, respectively. In the FFT-FBMC system, ${{\mu }_{\text{FFT-FBMC}}}=11.1\%,\text{ 5}\text{.9 }\!\!\%\!\!\text{ }$ and $\text{3 }\!\!\%\!\!\text{ }$ when $N=16,\,\,3\text{2}$ and $\text{64}$, respectively (according to the procedure of the FFT-FBMC, $N$ and ${\upsilon }'$ are replaced, instead of $L$ and $\upsilon $, to find $\mu $ also when the IOTA filter is used, the standard CP length is ${\upsilon }'=2$). Note that in the FBMC/OULP system ${{\mu }_{\text{FBMC/OULP}}}=0$ due to not using CP.

Fig. 8 depicts the BER performances of the OFDM, FFT-FBMC and FBMC/OULP systems versus the energy per bit (${{{E}}_{\text{b}}}\text{/}{{{N}}_{\text{0}}}$) for the SM-SISO channel. Since the results illustrated in this figure are achieved when the channel is SM, it can be assumed that the channel is constant during passing a frame of symbols. Thus, OFDM achieves the best performance that is independent of $L$. However, the performances of the FFT-FBMC and FBMC/OULP are improved by increasing $L$; where for $L=512$ both systems achieve a performance very close to that of the OFDM. These results come from the fact that for small values of $L$ in the FBMC-based systems, the channel spectrum is not flat enough over each subchannel (whereas the subchannel flatness is assumed in \eqref{22}); thus, it causes more BERs. By increasing $L$, channel spectrum would be flat over each subchannel that causes \eqref{22} to become valid; as a result, the performance of the FBMC/OULP improves. There is the similar situation for the FFT-FBMC system, as well.

The BER performances of the OFDM, FFT-FBMC and FBMC/OULP systems, when Alamouti coding scheme is used, are shown in Fig. 9 for the SM-MIMO channel with ${{N}_{\text{t}}}={{N}_{\text{r}}}=2$. As it is seen, similar to the SM-SISO channel, by increasing $L$ in the SM-MIMO channel the BERs of the FBMC/OULP and FFT-FBMC systems reach that of the OFDM. Fig. 10 and Fig. 11, respectively, indicate the performances in the FM-MIMO and VFM-MIMO channels when the Alamouti coding scheme is used. The results reveal that, in addition to its more bandwidth efficiency, the proposed FBMC/OULP system is less sensitive to time variations of the channel and outperforms FFT-FBMC and also OFDM (under suitable value of $L$). 

\begin{figure}[!t]
\centering
\includegraphics [width=3in]{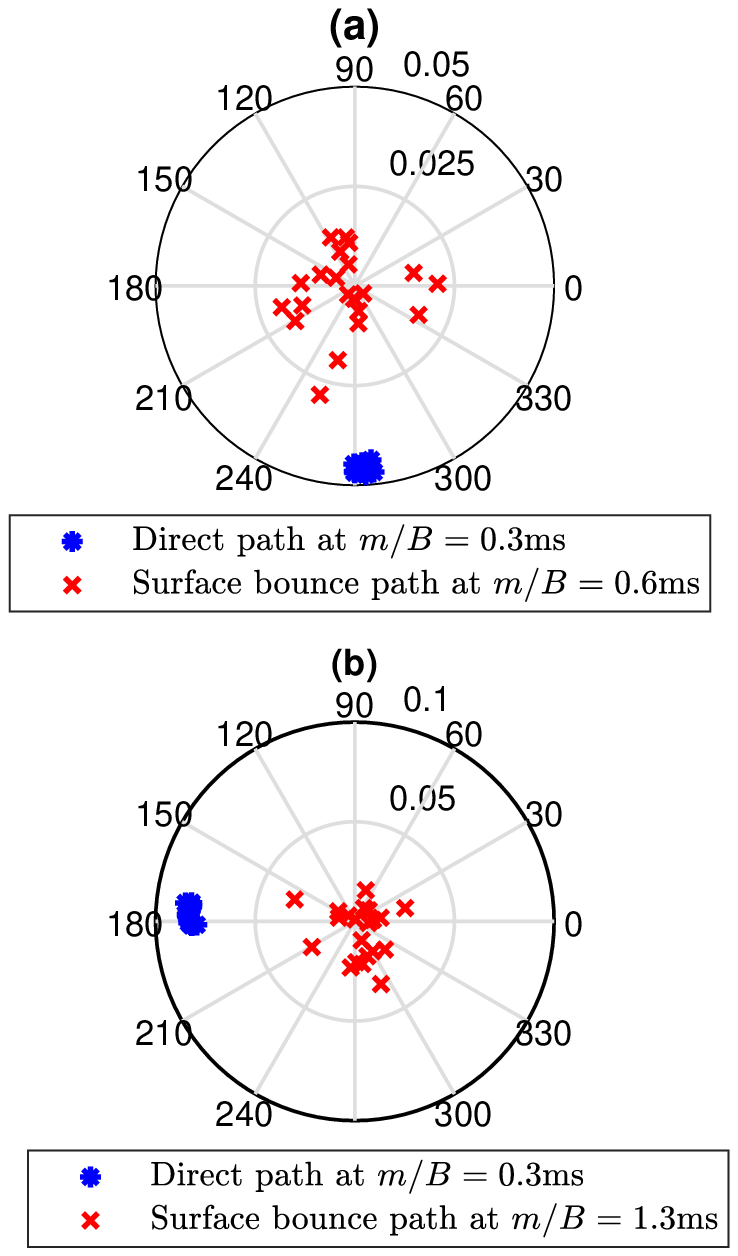}
\vspace{-0mm}
\caption{BER of the proposed FBMC/OULP, OFDM and FFT-FBMC in SM-SISO channel.} \vspace{-3mm}
 \label{F8}
\end{figure}
\begin{figure}[!t]
\centering
\includegraphics [width=3in]{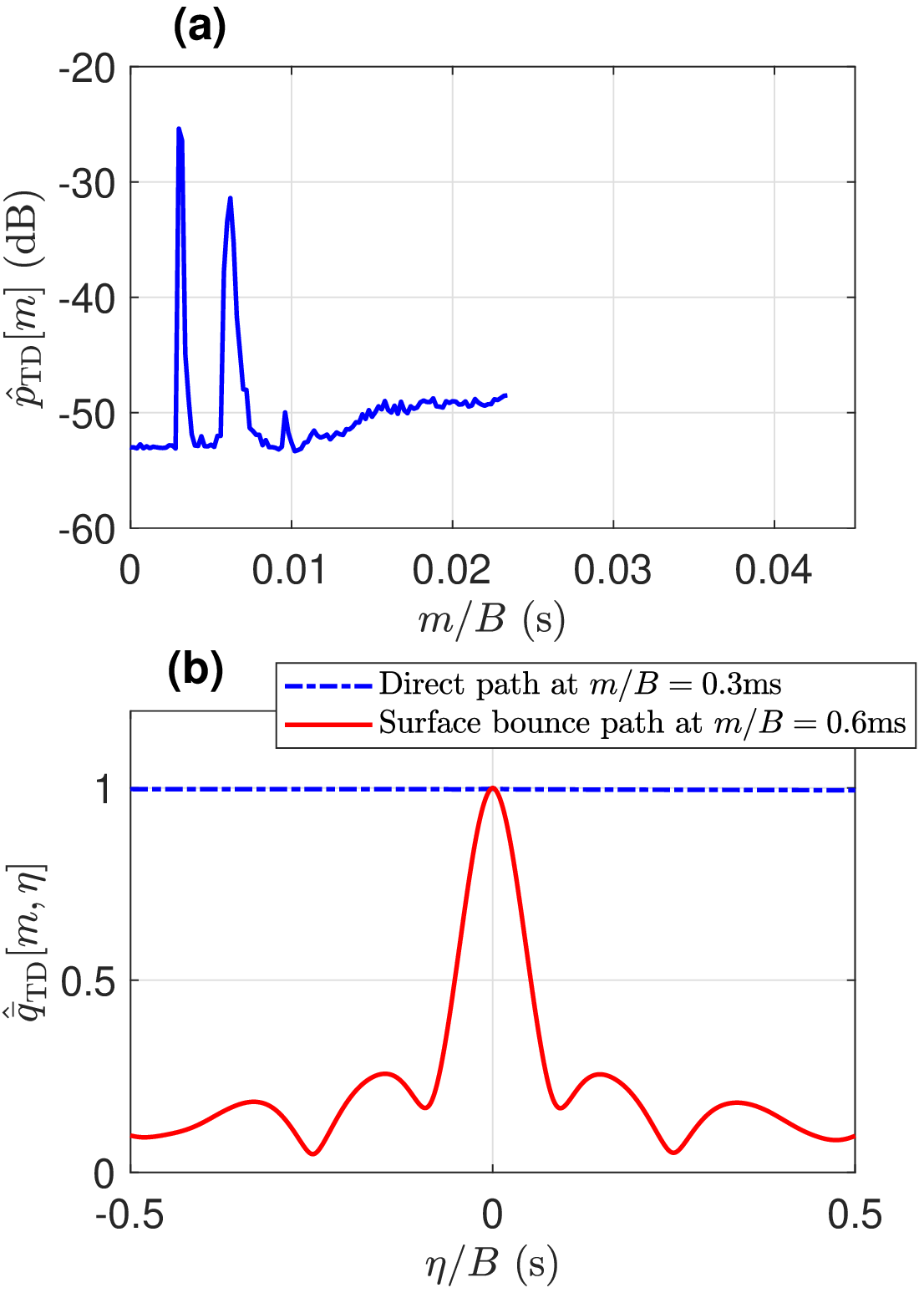}
\vspace{-0mm}
\caption{BER of the proposed FBMC/OULP, OFDM and FFT-FBMC in $2\times 2$ SM-MIMO channel when Alamouti coding is used.} 
\vspace{-3mm}
 \label{F9}
\end{figure}
\begin{figure}[!t]
\centering
\includegraphics [width=3in]{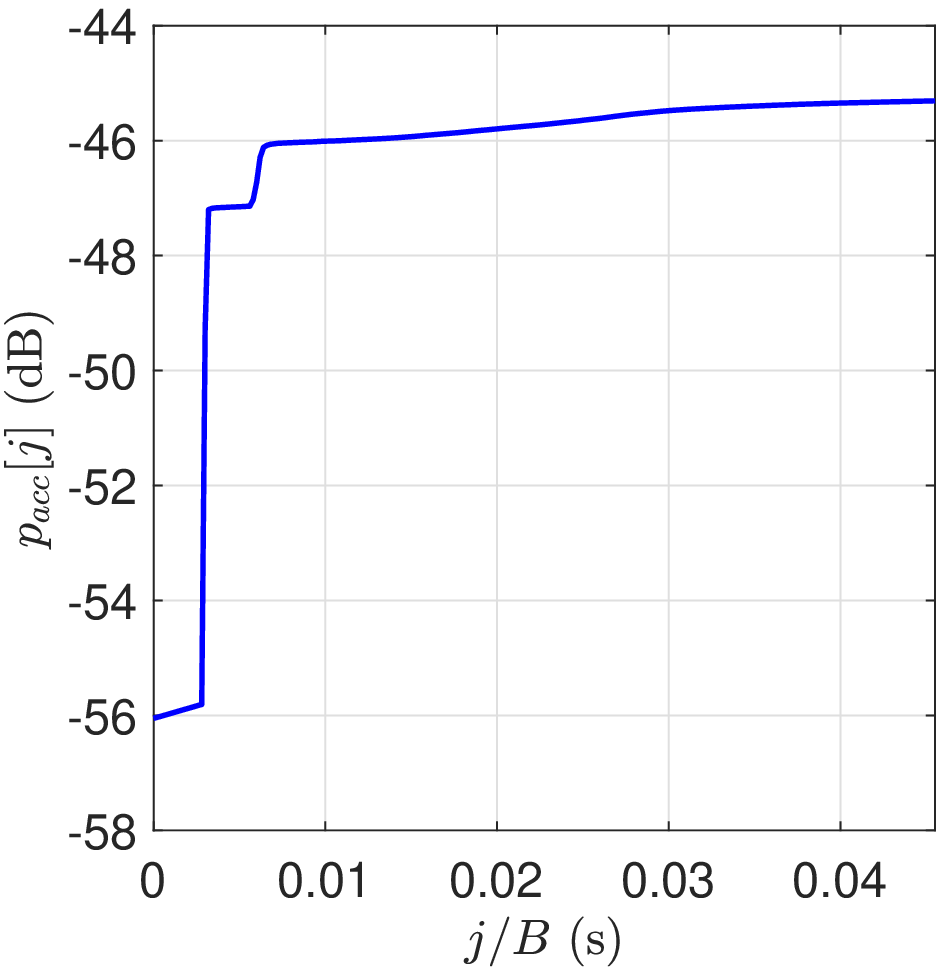}
\vspace{-0mm}
\caption{BER of the proposed FBMC/OULP, OFDM and FFT-FBMC in $2\times 2$ FM-MIMO channel when Alamouti coding is used.} 
\vspace{-3mm}
 \label{F10}
\end{figure}
\begin{figure}[!t]
\centering
\includegraphics [width=3in]{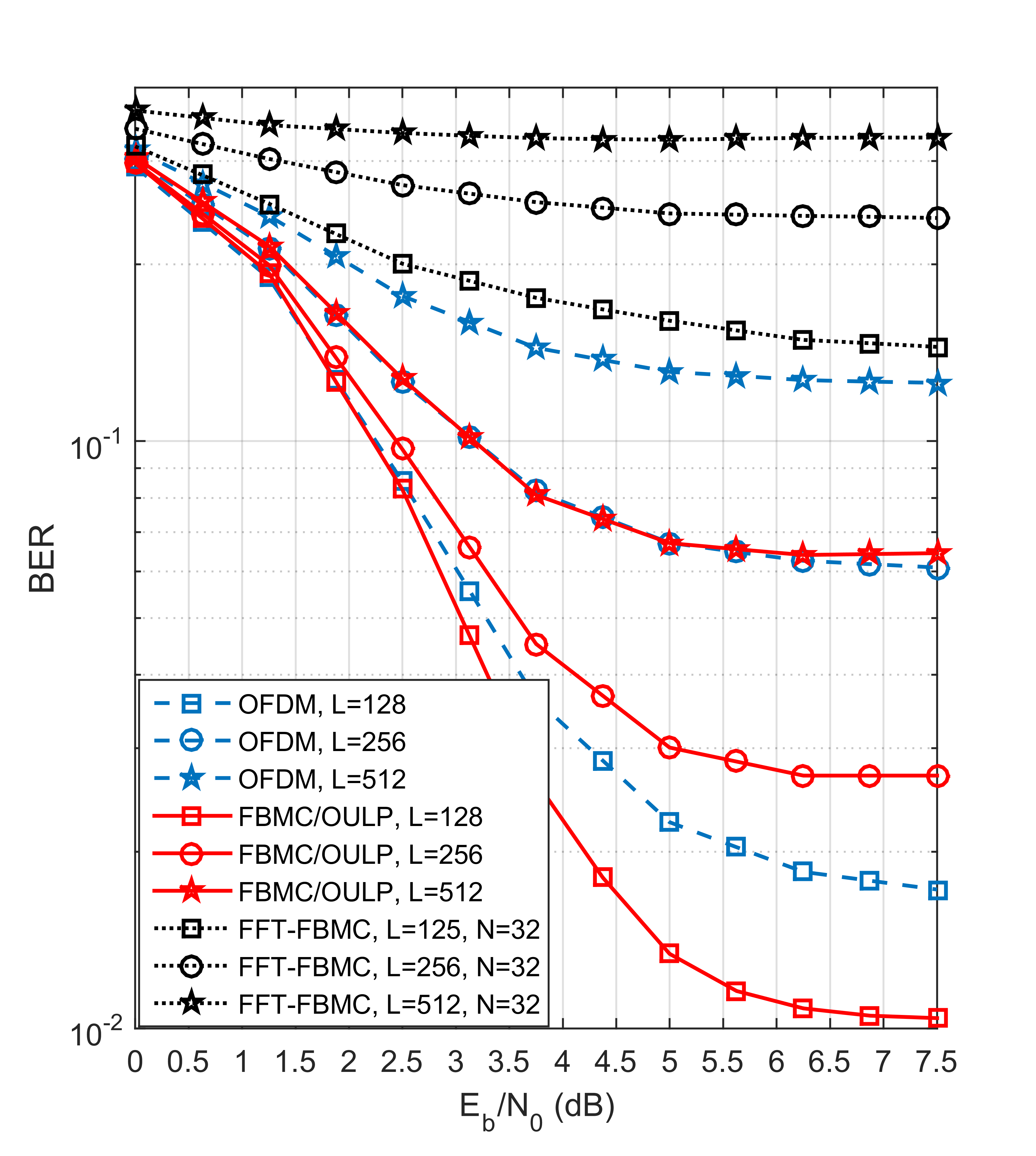}
\vspace{-0mm}
\caption{BER of the proposed FBMC/OULP, OFDM and FFT-FBMC for $2\times 2$ VFM-MIMO channel when Alamouti coding is used.} 
\vspace{-3mm}
 \label{F11}
\end{figure}

To evaluate the sensitivity of the FBMC/OULP system to time variation of channel and also compare it with those of the FFT-FBMC and OFDM systems, the BER performances of these systems for different coherence times at ${{E}_{b}}/{{N}_{0}}=7.5$ dB (according to Table II, for 16-QAM modulation it is equal to SNR=30 dB) are shown in Fig. 12, Fig. 13 and Fig. 14 for $L=128,\,\,256$ and $512$, respectively. These results indicate that for SM channel (about ${{T}_{\,\text{co}}}>800\,\,\text{ }\!\!\mu\!\!\text{ s}$), the FBMC/OULP system achieves the performance that comes to be competitive to the OFDM system performance when $L=512$. This result is achieved due to validity of \eqref{22} when the number of subchannels is large \cite{R29}. For the FM and VFM channels (about ${{T}_{\,\text{co}}}<800\,\,\text{ }\!\!\mu\!\!\text{ s}$), in comparison with the FFT-FBMC and OFDM systems, the FBMC/OULP system achieves a better performance when $L=128$. These results show that the OFDM and FFT-FBMC systems are very sensitive to channel time variation compared with the FBMC/OULP system. It should be noted that the symbol duration of the FFT-FBMC system is $L\,(N+{\upsilon }'+1)\,{{t}_{\text{s}}}/2$ which is very longer than that of the FBMC/OULP system (which is $\,L{{t}_{\text{s}}}$). This long duration leads to damaged symbols passing through the FM and VFM channels and significantly reduces the performance of the FFT-FBMC system.

To compare the BER performance and bandwidth efficiency of two filter-bank based systems, the FBMC/OULP and the FFT-FBMC, one can conclude that (by selecting suitable number of subchannels for both systems): \textit{i)} the performance of the FBMC/OULP system is very close to that of the FFT-FBMC system in the SM channels; \textit{ii)} the FBMC/OULP system significantly outperforms the FFT-FBMC system in the FM and VFM channels; \textit{iii)} because of using no CP, the FBMC/OULP system is more bandwidth efficient than the FFT-FBMC system. 

Moreover, it is remarkable that the FBMC/OULP system (similar to the other FBMC-based systems) has low flexibility in tuning the subchannel bandwidth and carrier frequencies when it is implemented by polyphase networks \cite{R41}. This issue becomes more serious in some scenarios such as fragmented spectrum, cognitive radio and dynamic spectrum allocation. Nevertheless, an alternative method has been proposed in \cite{R42} to implement the FBMC systems which properly tackles the mentioned issue. Furthermore, it should be noted that the FBMC/OULP system can be used in multi-user with shared spectrum scenario in which the circular convolutional property is sustained.

\begin{figure}[!t]
\centering
\includegraphics [width=3in]{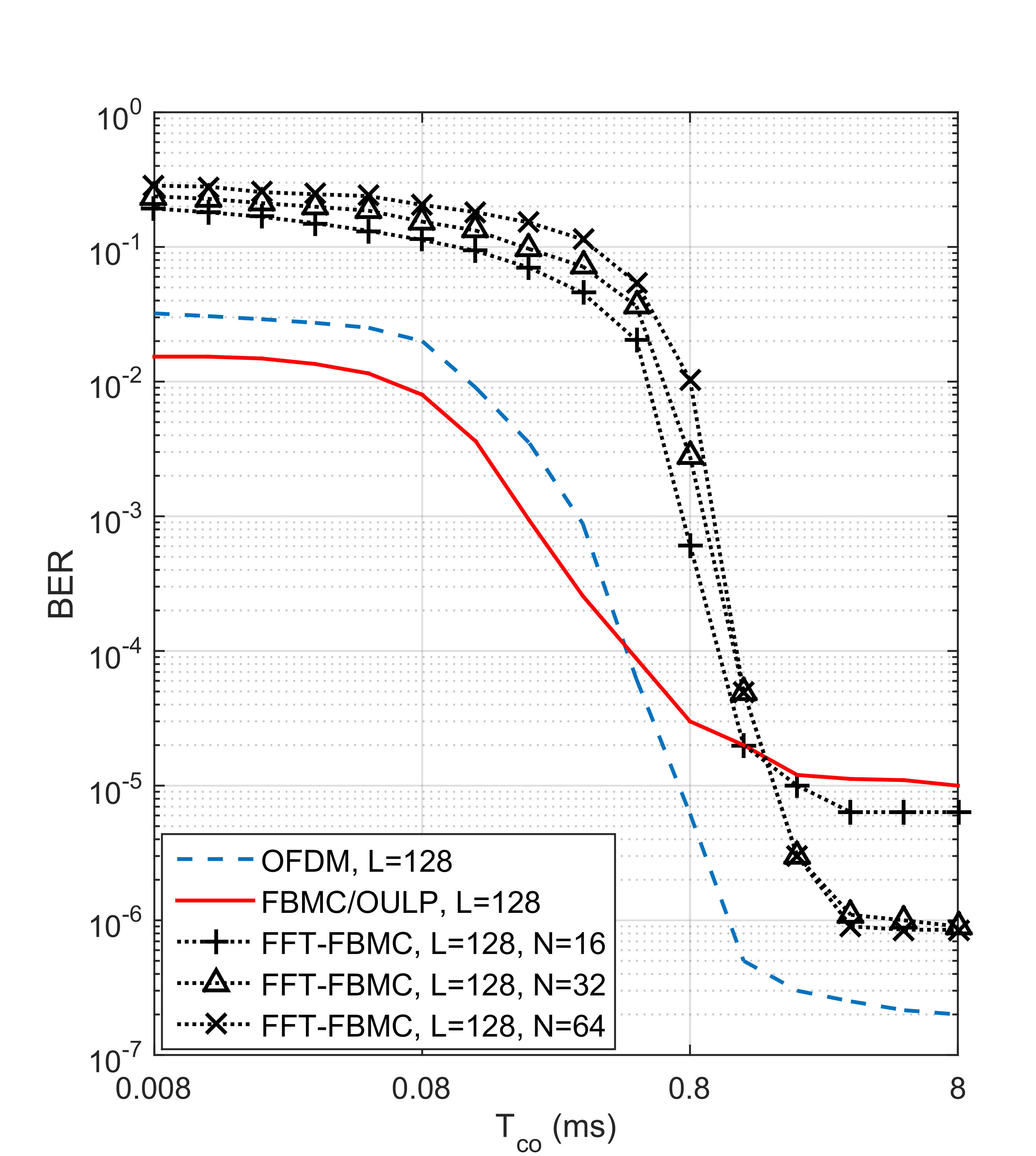}
\vspace{-0mm}
\caption{BER versus channel coherence time for the proposed FBMC/OULP, OFDM and FFT-FBMC in $2\times 2$ MIMO channel when $L=128$, ${{E}_{b}}/{{N}_{0}}=7.5$dB and Alamouti coding is used.} 
\vspace{-3mm}
 \label{F12}
\end{figure}
\begin{figure}[!t]
\centering
\includegraphics [width=3in]{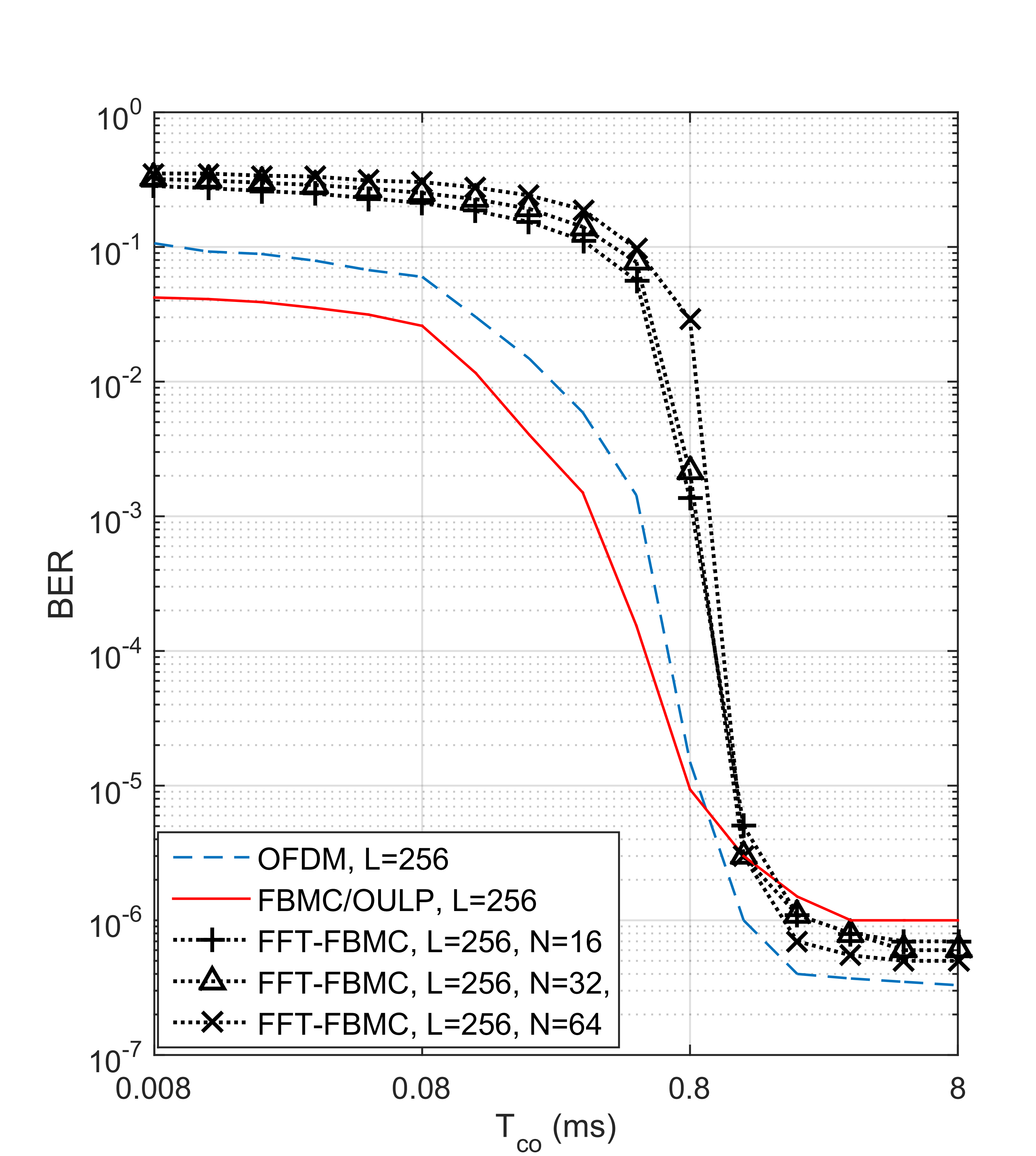}
\vspace{-0mm}
\caption{BER versus channel coherence time for the proposed FBMC/OULP, OFDM and FFT-FBMC in $2\times 2$ MIMO channel when $L=256$, ${{E}_{b}}/{{N}_{0}}=7.5$dB and Alamouti coding is used.}
\vspace{-3mm}
 \label{F13}
\end{figure}
\begin{figure}[!t]
\centering
\includegraphics [width=3in]{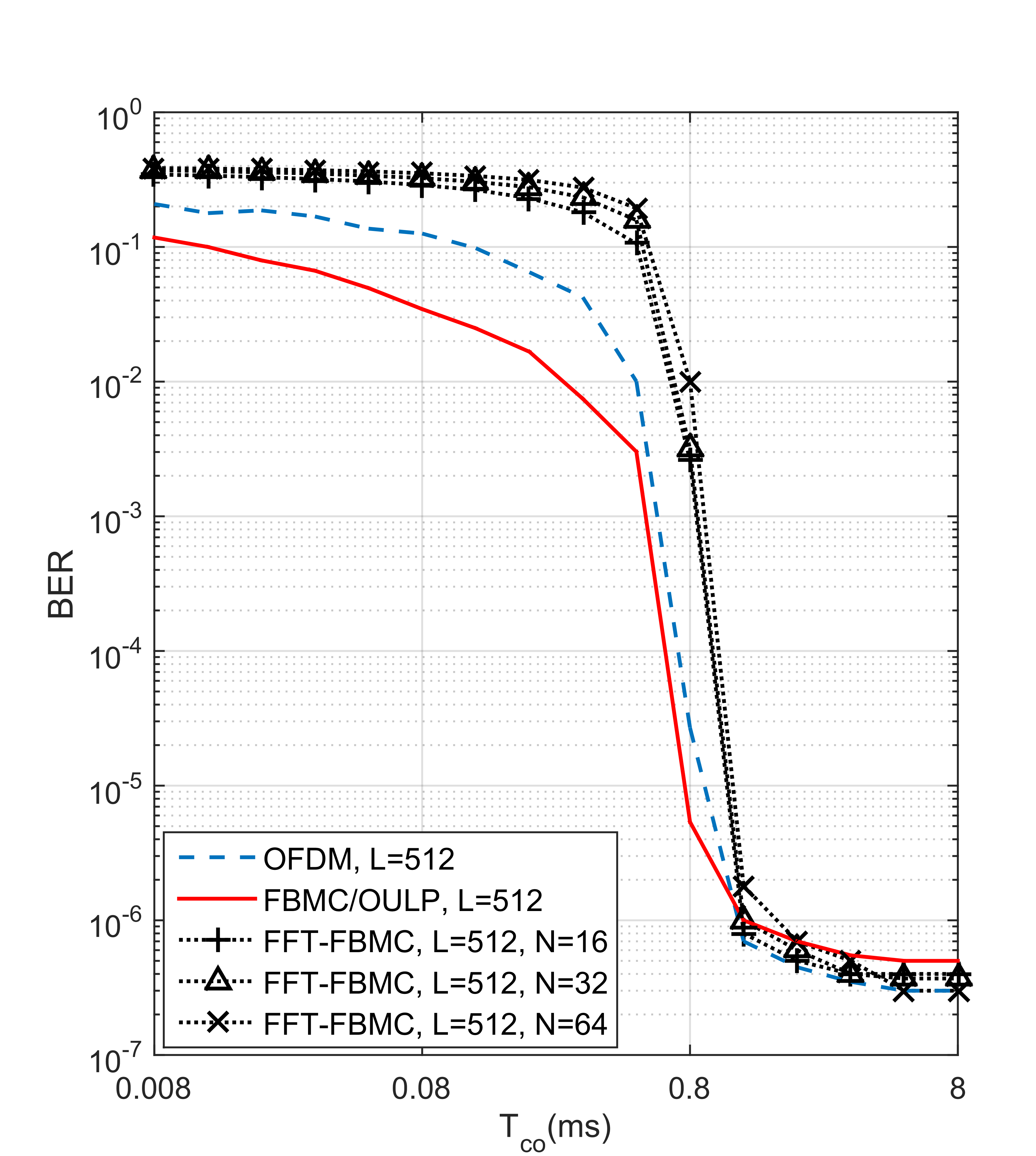}
\vspace{-0mm}
\caption{BER versus channel coherence time for the proposed FBMC/OULP, OFDM and FFT-FBMC in $2\times 2$ MIMO channel when $L=512$, ${{E}_{b}}/{{N}_{0}}=7.5$dB and Alamouti coding is used.}
\vspace{-3mm}
 \label{F14}
\end{figure}

\section{Conclusion}
FBMC systems suffer from an intrinsic interference when complex-valued symbols are transmitter over the time and frequency domains with the data symbol density of one. This challenging issue limits the full diversity gain achievement in MIMO FBMC systems when OSTBC codes are used.  A new filter-bank based system, called FBMC/OULP, has been proposed in this paper to mitigate the intrinsic interference. In the proposed FBMC/OULP system, complex-valued symbols are loaded on the upper and lower branches of the polyphase network alternatively with offset time such that the symbol density of the time-frequency lattice becomes equal to one complex-valued symbol. The intrinsic interference becomes zero with a good approximation in the proposed FBMC/OULP system when the channel is frequency flat and a well-localized time-frequency prototype filter is used.

In frequency selective channels, the produced interference in the FBMC/OULP system can be eliminated by using the MMSE estimator by using the circular convolution property of interfered and desired parts of the received signal. Due to using complex-valued symbols and also diminishing the interference, the FBMC/OULP system is potentially able to achieve the full diversity gain of the MIMO channels by employing a low complexity ML detection of OSTBC. Simulation results have indicated that by selecting a suitable number of subchannels, the proposed FBMC/OULP system, in addition to its bandwidth efficiency, outperforms the OFDM system in fast multi-path fading channels. Also, the performance of the FBMC/OULP system is competitive with that of the OFDM in slow multi-path fading channels.

\appendices
\section{Proof of \eqref{11}}

We have $\mathbf{V}_{\kappa}^{{{k}'}}=\mathbf{F}_{L}^{\dagger }\mathbf{Z}_{k}^{{{k}'}}{{\mathbf{F}}_{L}}$; thus, according to \eqref{8}, the $(a,b)\text{th}$ element of $\mathbf{V}_{\kappa}^{{{k}'}}$ can be written as
\begin{equation}
\mathbf{V}_{\kappa}^{{{k}'}}(a,b)=\sum\limits_{\ell =-\Delta }^{\Delta }{(\xi _{\kappa,\ell }^{{{k}'}}\,{{e}^{j2\pi a\ell /L}}\,\sum\limits_{c=0}^{L-1}{{{e}^{j2\pi c(a-b)/L}})}}.
\label{43}
\end{equation}
As it can be seen, $\mathbf{V}_{\kappa}^{{{k}'}}(a,b)=0$, for $a\ne b$ and $\mathbf{V}_{\kappa}^{{{k}'}}(a,b)=\sum\nolimits_{\ell =-\Delta }^{\Delta }{\xi _{k,0}^{{{k}'}}\,{{e}^{j2\pi a\ell /L}}}$, for $a=b$. Consequently, $\mathbf{V}_{\kappa}^{{{k}'}}$ is a diagonal matrix such that $\mathbf{V}_{\kappa}^{{{k}'}}=\text{diag(}\mathbf{F}_{L}^{\dagger }\bar{\xi }_{k}^{{{k}'}})$ \cite{R43}.

\section{Proof of \eqref{25}}
According to \eqref{25} we have ${{\mathbf{U}}_{{{k}'}}}=\mathbf{F}_{L}^{\dagger }{{\mathbf{H}}_{{{k}'}}}\,{{\mathbf{F}}_{L}}$. Thus, if we define $\mathbf{A}\triangleq {{\mathbf{F}}_{L}}{{\mathbf{U}}_{{{k}'}}}\mathbf{F}_{L}^{\dagger }$ and show that $\mathbf{A}={{\mathbf{H}}_{{{k}'}}}$, then \eqref{25} is proved. In this regard, the $(a,b)\text{th}$ entry of $\mathbf{A}$ can be written as
\begin{equation}
\mathbf{A}(a,b)=\sum\limits_{m=0}^{{{L}_{\text{C}}}-1}{{{h}_{{k}',m}}\,{{e}^{-j2\pi am/L}}\,\sum\limits_{c=0}^{L-1}{{{e}^{-j2\pi c(a-b)/L}}}}.
\label{44}
\end{equation}
As it can be seen, $\mathbf{A}(a,b)=0$, for $a\ne b$ and $\mathbf{A}(a,b)=\sum\nolimits_{m=0}^{{{L}_{\text{C}}}-1}{{{h}_{{k}',m}}\,{{e}^{-j2\pi a\ell /L}}}$, for $a=b$. Consequently, referring to \eqref{23}, it is obvious that $\mathbf{A}={{\mathbf{H}}_{{{k}'}}}$. Accordingly it is concluded that ${{\mathbf{H}}_{{{k}'}}}={{\mathbf{F}}_{L}}{{\mathbf{U}}_{{{k}'}}}\mathbf{F}_{L}^{\dagger }$ then it yields to ${{\mathbf{U}}_{{{k}'}}}=\mathbf{F}_{L}^{\dagger }{{\mathbf{H}}_{{{k}'}}}\,{{\mathbf{F}}_{L}}$.

\section{Proof of \eqref{35}}
Considering \eqref{32}, we have ${{\tilde{\bar{R}}}_{{{k}'}}}={{\mathbf{U}}_{{{k}'}}}\bar{Y}_{{{k}'}}^{\rm{De}}+{{\bar{\Omega }}_{{{k}'}}}$. By considering \eqref{17} and \eqref{25}, for an even ${k}'$, \eqref{32} can be expanded as
\begin{equation}
\begin{array}{l}
{\left[ {\begin{array}{*{20}{c}}
{{{\tilde R}_{k',0}}}\\
 \vdots \\
{{{\tilde R}_{k',L - 1}}}
\end{array}} \right]_{L \times \,1}} = \\
\\
{\left[ {\begin{array}{*{20}{c}}
{{h_{k',0}}}&{\overbrace {\bf{0}}^{L - {L_{\rm{c}}}}}&{{h_{k',{L_{\rm{c}}} - 1}}}& \cdots &{{h_{k',1}}}\\
 \vdots & \searrow &{\bf{0}}& \searrow & \vdots \\
{{h_{k',{L_{\rm{c}}} - 1}}}&{}&{{h_{k',0}}}&{}&{{h_{k',{L_{\rm{c}}} - 1}}}\\
{}& \searrow & \vdots & \searrow &{\bf{0}}\\
{\bf{0}}&{}&{{h_{k',{L_{\rm{c}}} - 1}}}& \cdots &{{h_{k',0}}}
\end{array}} \right]_{L \times L}}\\
 \times \left[ {\begin{array}{*{20}{c}}
{{q_{k',0}}}& \cdots &{{q_{k',L/2 - 1}}}&0& \cdots &0
\end{array}} \right]_{\,\,L \times \,1}^{\,T}\\
 + \left[ {\begin{array}{*{20}{c}}
{{\Omega _{k',0}}}& \cdots &{{\Omega _{k',L - 1}}}
\end{array}} \right]_{\,L \times \,1\,}^{\,T}.
\end{array}
\label{45}
\end{equation}
On the other hand, from \eqref{33}, we have ${{\bar{x}}_{{{k}'}}}=\mathbf{G}_{{{k}'}}^{{{L}_{\text{c}}}}\,{{\tilde{\bar{R}}}_{{{k}'}}}$. Considering the structure of $\mathbf{G}_{{{k}'}}^{{{L}_{\text{c}}}}$ in \eqref{34} it leads to
\begin{equation}
\begin{array}{l}
{\left[ {\begin{array}{*{20}{c}}
{{x_{k',0}}}\\
 \vdots \\
{{x_{k',L/2 - 1}}}
\end{array}} \right]_{L/2 \times \,1}} = \\
\\
{\left[ {\begin{array}{*{20}{c}}
{{h_{k',0}}}&{\overbrace {\bf{0}}^{L/2 - {L_{\rm{c}}}}}&{{h_{k',{L_{\rm{c}}} - 1}}}& \cdots &{{h_{k',1}}}\\
 \vdots & \searrow &{\bf{0}}& \searrow & \vdots \\
{{h_{k',{L_{\rm{c}}} - 1}}}&{}&{{h_{k',0}}}&{}&{{h_{k',{L_{\rm{c}}} - 1}}}\\
{}& \searrow & \vdots & \searrow &{\bf{0}}\\
{\bf{0}}&{}&{{h_{k',{L_{\rm{c}}} - 1}}}& \cdots &{{h_{k',0}}}
\end{array}} \right]_{\,L/2 \times Ll2}}\\
 \times \left[ {\begin{array}{*{20}{c}}
{{q_{k',0}}}& \cdots &{{q_{k',L/2 - 1}}}
\end{array}} \right]_{\,\,L/2 \times \,1}^{\,T}\\
 + \left[ {\begin{array}{*{20}{c}}
{{\psi _{k',0}}}& \cdots &{{\psi _{k',L/2 - 1}}}
\end{array}} \right]_{\,L/2 \times \,1\,}^{\,T};
\end{array}
\label{46}
\end{equation}
As is obvious, we can write ${{\bar{x}}_{{{k}'}}}={{\mathbf{u}}_{{{k}'}}}{{\bar{q}}_{{{k}'}}}+{{\bar{\psi }}_{{{k}'}}}$ in which ${{\mathbf{u}}_{{{k}'}}}$ is a circular $(L/2)\times (L/2)$ matrix of time domain CIR coefficients. For an odd ${k}'$, this equation can be derived in the same way.

\section{Proof of \eqref{38}}

Since ${{\mathbf{h}}_{{{k}'}}}={{\mathbf{F}}_{L/2}}{{\mathbf{u}}_{{{k}'}}}\mathbf{F}_{L/2}^{\dagger }$, by considering \eqref{36}, the $(a,b)\text{th}$ element of  ${{\mathbf{h}}_{{{k}'}}}$ yields to
\begin{equation}
{{\mathbf{h}}_{{{k}'}}}(a,b)=\sum\limits_{m=0}^{{{L}_{\text{C}}}-1}{{{h}_{{k}',m}}\,{{e}^{-j4\pi am/L}}\,\sum\limits_{c=0}^{L/2-1}{{{e}^{-j4\pi c(a-b)/L}}}}.
\label{47}
\end{equation}
As it can be seen, ${{\mathbf{h}}_{{{k}'}}}(a,b)=0$, for $a\ne b$ and ${{\mathbf{h}}_{{{k}'}}}(a,b)=\sum\nolimits_{m=0}^{{{L}_{\text{C}}}-1}{{{h}_{{k}',m}}\,{{e}^{-j4\pi a\ell /L}}}$, for $a=b$. As a result, ${{\mathbf{h}}_{{{k}'}}}=\text{diag}({{\mathbf{F}}_{L/2}}{{\bar{h}}_{{{k}'}}})$.

\section{Time domain correlation of the output noise}

According to \eqref{3}, the trapped noise at the receiver of the FBMC/OULP, at ${k}'\text{th}$ time slot and ${l}'\text{th}$ subchannel is ${{\omega }_{{k}',{l}'}}=\sum\nolimits_{m=-\infty }^{\infty }{w\,(m)\,f_{{k}',{l}'}^{*}(m)}$. Thus, the time-frequency correlation of ${{\omega }_{{k}',{l}'}}$ is \cite{R44}
\begin{equation}
\begin{array}{l}
{\rm{E}}\{ {\omega _{k',l'}}{\mkern 1mu} \omega _{k' - k,l' - \ell }^*\}  = \\
\,\,\,\,\,\,\,\,\,\,\,\,\,{N_0}{\mkern 1mu} \sum\limits_{m =  - \infty }^\infty  {f_{k',l'}^*(m){\mkern 1mu} {\mkern 1mu} {f_{k' - k,l' - \ell }}(m)}  = {N_0}{\mkern 1mu} \xi _{k,\ell }^{k'},
\end{array}
\label{48}
\end{equation}
where $\text{E}\{.\}$ presents the mathematical expectation operator. As \eqref{48} depicts, ${{\omega }_{{k}',{l}'}}$ is a colored noise within the time and frequency axes. On the other hand, according to \eqref{24}, ${{\bar{\Omega }}_{{{k}'}}}=\mathbf{F}_{L}^{\dagger }{{\bar{\omega }}_{{{k}'}}}$. Thus, the 2-dimention correlation of ${{\Omega }_{{k}',n}}$ (where ${{\bar{\Omega }}_{{{k}'}}}\triangleq [{{\Omega }_{{k}',0}},{{\Omega }_{{k}',1}},\ldots $ ${{\Omega }_{{k}',L-1}}{{]}^{T}}$), across ${k}'$ and $n$, becomes
\begin{equation}
\begin{array}{l}
{\rm{E}}\{ {\Omega _{k',n}}{\mkern 1mu} \Omega _{k' - \kappa,n - {n_1}}^*\}  = \\
\,\,\,\,\,\,\,\,\,\,\,\frac{1}{L}\sum\limits_{a = 0}^{L - 1} {\sum\limits_{b = 0}^{L - 1} {{\mkern 1mu} {\rm{E}}} } \left\{ {{\omega _{k',a}}{\mkern 1mu} \omega _{k' - \kappa,b}^*} \right\}{e^{j2\pi na/L}}{\mkern 1mu} {e^{ - j2\pi (n - {n_1}){\kern 1pt} b/L}} = \\
{\mkern 1mu} {\mkern 1mu} {\mkern 1mu} {\mkern 1mu} {\mkern 1mu} {\mkern 1mu} {\mkern 1mu} {\mkern 1mu} {\mkern 1mu} {\mkern 1mu} {\mkern 1mu} {\mkern 1mu} {\mkern 1mu} {\mkern 1mu} {\mkern 1mu} {\mkern 1mu} {\mkern 1mu} {\mkern 1mu} {\mkern 1mu} {\mkern 1mu} {\mkern 1mu} {\mkern 1mu} {\mkern 1mu} {\mkern 1mu} {\mkern 1mu} {\mkern 1mu} \,\,\frac{{{N_0}}}{L}\sum\limits_{a = 0}^{L - 1} {\sum\limits_{b = 0}^{L - 1} {{\mkern 1mu} {\mkern 1mu} \xi _{\kappa,a - b}^{k'}{\mkern 1mu} } } {e^{j2\pi n{\kern 1pt} a/L}}{\mkern 1mu} {e^{ - j2\pi (n - {n_1}){\kern 1pt} b/L}} = \\
{\mkern 1mu} {\mkern 1mu} {\mkern 1mu} {\mkern 1mu} {\mkern 1mu} {\mkern 1mu} {\mkern 1mu} {\mkern 1mu} {\mkern 1mu} {\mkern 1mu} \,\,\,\,\,\,\,\frac{{{N_0}}}{L}\sum\limits_{b = 0}^{L - 1} {{\rm Z}_{\kappa,n}^{k'}{\mkern 1mu} {e^{j2\pi {\kern 1pt} {n_1}b{\kern 1pt} /L}} = {N_0}{\mkern 1mu} {\mkern 1mu} } {\rm Z}_{\kappa,n}^{k'}{\mkern 1mu} {\delta _{{n_1}}}.
\end{array}
\label{49}
\end{equation}

This result illustrates that ${{\Omega }_{{k}',n}}$ elements are independent regarding the index $n$  and are correlated across the index ${k}'$. In addition, from \eqref{33}, we have ${{\bar{\psi }}_{{{k}'}}}=\mathbf{G}_{{{k}'}}^{{{L}_{\text{c}}}}\,{{\bar{\Omega }}_{{{k}'}}}$ and also from \eqref{37}, ${{\bar{\Psi }}_{{{k}'}}}={{\mathbf{F}}_{L/2}}{{\bar{\psi }}_{{{k}'}}}$; as a result, ${{\bar{\Psi }}_{{{k}'}}}={{\mathbf{F}}_{L/2}}\mathbf{G}_{{{k}'}}^{{{L}_{\text{c}}}}\,{{\bar{\Omega }}_{{{k}'}}}$. Since, in multicarrier systems, always ${{L}_{\text{c}}}\ll L$, it can be approximated that $\mathbf{G}_{{{k}'}}^{{{L}_{\text{c}}}}\approx \mathbf{G}_{{{k}'}}^{1}$ (refer to \eqref{19} and \eqref{34}). Consequently, it yields to
\begin{equation}
{\bar \Psi _{k'}} = {{\bf{F}}_{L/2}}{\bf{G}}_{k'}^1{\mkern 1mu} {\bar \Omega _{k'}}.
\label{50}
\end{equation}
Accordingly, the time domain correlation of ${{\Psi }_{{k}',n}}$ elements (across the index ${k}'$), for an even ${k}'$, can be written as \cite{R45}
\begin{equation}
\begin{array}{l}
{\phi _k} \buildrel \Delta \over = {\rm{E}}\{ {\Psi _{k',n}}{\mkern 1mu} \Psi _{k' - \kappa,n{\kern 1pt} }^*\}  = {\mkern 1mu} {\mkern 1mu} \\
\,\,\,\,\,\,\,\,\,\frac{2}{L}\sum\limits_{a = 0}^{L/2 - 1} {\sum\limits_{b = 0}^{L/2 - 1} {{\mkern 1mu} {\rm{E}}\{ } } {\Omega _{k',a}}{\mkern 1mu} \Omega _{k' - \kappa,b}^*\} {\mkern 1mu} {e^{ - j4\pi n{\kern 1pt} a/L}}{\mkern 1mu} {e^{j4\pi n{\kern 1pt} b/L}} = \\
{\mkern 1mu} {\mkern 1mu} {\mkern 1mu} {\mkern 1mu} {\mkern 1mu} {\mkern 1mu} {\mkern 1mu} {\mkern 1mu} {\mkern 1mu} {\mkern 1mu} {\mkern 1mu} {\mkern 1mu} {\mkern 1mu} {\mkern 1mu} {\mkern 1mu} {\mkern 1mu} {\mkern 1mu} {\mkern 1mu} {\mkern 1mu} \,\, \frac{{2{N_0}}}{L}\sum\limits_{a = 0}^{L/2 - 1} {{\rm Z}_{\kappa,a}^{k'}{\mkern 1mu} } .
\end{array}
\label{51}
\end{equation}

 In the same way, when ${k}'$ is an odd number, the time correlation is ${\phi _\kappa} = 2{N_0}/L \times \sum\nolimits_{a = L/2}^{L - 1} {{\rm Z}_{\kappa,a}^{k'}} $, which according to \eqref{12}, is exactly equal to the time correlation when ${k}'$ is an even number. Since ${{\phi }_{k}}$ just depends on $k$ and is free from ${k}'$ and $n$, the time domain correlation of the FBMC/OULP output noise, in each subcarrier, is stationary and free from the index of the subchannel. Thus, in the case of the IOTA as the prototype filter, the normalized time domain correlation is ${{\phi }_{\,k}}/{{\phi }_{\,0}}<0.3$ when $\left| k \right|>0$.

% Can use something like this to put references on a page
% by themselves when using endfloat and the captionsoff option.
\ifCLASSOPTIONcaptionsoff
  \newpage
\fi

% trigger a \newpage just before the given reference
% number - used to balance the columns on the last page
% adjust value as needed - may need to be readjusted if
% the document is modified later
%\IEEEtriggeratref{8}
% The "triggered" command can be changed if desired:
%\IEEEtriggercmd{\enlargethispage{-5in}}

% references section

% can use a bibliography generated by BibTeX as a .bbl file
% BibTeX documentation can be easily obtained at:
% http://mirror.ctan.org/biblio/bibtex/contrib/doc/
% The IEEEtran BibTeX style support page is at:
% http://www.michaelshell.org/tex/ieeetran/bibtex/
%\bibliographystyle{IEEEtran}
% argument is your BibTeX string definitions and bibliography database(s)
%\bibliography{IEEEabrv,../bib/paper}
%
% <OR> manually copy in the resultant .bbl file
% set second argument of \begin to the number of references
% (used to reserve space for the reference number labels box)
\bibliographystyle{IEEEtranTCOM}

\bibliography{Refrences}

% biography section
% 
% If you have an EPS/PDF photo (graphicx package needed) extra braces are
% needed around the contents of the optional argument to biography to prevent
% the LaTeX parser from getting confused when it sees the complicated
% \includegraphics command within an optional argument. (You could create
% your own custom macro containing the \includegraphics command to make things
% simpler here.)
%\begin{IEEEbiography}[{\includegraphics[width=1in,height=1.25in,clip,keepaspectratio]{mshell}}]{Michael Shell}
% or if you just want to reserve a space for a photo:

% You can push biographies down or up by placing
% a \vfill before or after them. The appropriate
% use of \vfill depends on what kind of text is
% on the last page and whether or not the columns
% are being equalized.

%\vfill

% Can be used to pull up biographies so that the bottom of the last one
% is flush with the other column.
%\enlargethispage{-5in}

% that's all folks
\end{document}